\documentclass[fleqn,usenatbib]{mnras}

\usepackage{newtxtext,newtxmath}
\usepackage{url}

\usepackage[T1]{fontenc}
\usepackage[dvipsnames]{xcolor}

\usepackage[normalem]{ulem}

\DeclareRobustCommand{\VAN}[3]{#2}
\let\VANthebibliography\thebibliography
\def\thebibliography{\DeclareRobustCommand{\VAN}[3]{##3}\VANthebibliography}

\usepackage{graphicx}	
\usepackage{amsmath}	

\title[$\alpha$-bimodality in FIRE-2]{Effect of gas accretion on $\alpha$-element bimodality in Milky Way-mass galaxies in the FIRE-2 simulations}

\author[H. Parul et al.]{
Hanna Parul,$^{1,2}$\thanks{E-mail: hparul@crimson.ua.edu}
Jeremy Bailin,$^{1}$
Sarah R. Loebman,$^{3}$
Andrew Wetzel,$^{4}$
Megan Barry,$^{4}$
Binod Bhattarai$^{3}$
\\
$^{1}$Department of Physics and Astronomy, University of Alabama, Box 870324, Tuscaloosa, AL, 35487, USA\\
$^{2}$LIRA, Observatoire de Paris, Université PSL, Sorbonne Université, Université Paris Cité, CY Cergy Paris Université, CNRS, 92190 Meudon, France\\
$^{3}$Department of Physics, University of California, Merced, CA 95343, USA\\
$^{4}$Department of Physics \& Astronomy, University of California, Davis, CA 95616, USA\\
}

\date{Accepted XXX. Received YYY; in original form ZZZ}

\pubyear{2015}

\begin{document}
\label{firstpage}
\pagerange{\pageref{firstpage}--\pageref{lastpage}}
\maketitle

\begin{abstract}
We analyse the stellar distributions on the [Fe/H]-[Mg/Fe] plane for 11 Milky Way-mass galaxies from the FIRE-2 cosmological baryonic zoom-in simulations. Alpha-element bimodality, in the form of two separate sequences on the [Fe/H]-[Mg/Fe] plane, is not a universal feature of disk galaxies. Five galaxies demonstrate double sequences with the $\alpha$-enriched one being older and kinematically hotter, in qualitative agreement with the high-$\alpha$ and low-$\alpha$ populations in the Milky Way disk; three galaxies have unimodal distribution, two show weakly-bimodal features where low-$\alpha$ sequence is visible only over a short range of metallicities, and one show strong bimodality with a different slope of high-$\alpha$ population.
We examine the galaxies' gas accretion history over the last 8 Gyr, when bimodal sequences emerge, and demonstrate that the presence of the low-$\alpha$ sequence in the bimodal galaxies is related to the recent infall of metal-poor gas from the circumgalactic medium that joins the galaxy in the outskirts and induces significant growth of the gas disks compared to their non-bimodal counterparts. We also analyse the sources of the accreted gas and illustrate that both gas-rich mergers and smooth accretion of ambient gas can be the source of the accreted gas, and create slightly different bimodal patterns.
\end{abstract}




\section{Introduction}
The formation and evolution of the Milky Way are among the core questions of galactic astrophysics research today. Our location within the Galaxy provides an opportunity to study it with a level of detail impossible for any other galaxy. Vast amounts of data from modern surveys like APOGEE \citep{APOGEE}, Gaia \citep{Gaia}, GALAH \citep{galah}, and LAMOST \citep{lamost} allow for accurate mapping of the Milky Way's structure and facilitates uncovering its history.

For over 40 years it has been known that the galactic disk has a composite structure that can be described by two exponential components with different scale heights and scale lengths: a more centrally concentrated, vertically extended \textit{thick} disk and more horizontally extended but vertically compact \textit{thin} disk. The two disks also display differences in their vertical velocity dispersion, with the spatially selected thick disk containing dominantly kinematically hotter stellar orbits \citep{GilmoreReid83, soubiran2003, Kordopatis2013, Bensby2011}. 
Chemo-chronological analysis of the galactic disk also reveals the presence of multiple populations. 
One of the main features of the [Fe/H]-[$\alpha$/Fe] plane in the solar neighbourhood is the existence of high and low-$\alpha$ components that merge slightly above solar metallicity, which is known as $\alpha$-bimodality \citep{Fuhrmann98, Bensby2012, Adibekyan2013, Haywood2013, SilvaAguirre18, Vincenzo2021a}. Large spectroscopic surveys have revealed that the relative importance of each component varies with radius and vertical distance from the disk midplane \citep{Anders2014, Nidever2014, Hayden2015, Queiroz2019, Imig2023}. The high-$\alpha$ population is older, more compact, and prevails at higher $z$, and often is associated with the thick disk. The low-$\alpha$ sequence is younger, kinematically cooler and extends further to the outer disk, mostly overlaping with the thin disk  \citep[e.g][]{Nidever2014, Bensby2014}. Another feature of the low-$\alpha$ component is that the locus of metallicities depends on the radius, with the most metal-poor stars residing farther from the galaxy's center, reflecting the radial gradient of [Fe/H] \citep{Hayden2015}. However, it is still a matter of debate whether the two components are truly distinct and whether an exact mapping between geometrically and chemically-defined disks is possible \citep{Bovy2012, Hayden2017, Mackereth2017, Queiroz2023, BinneyVasiliev2024}. Surveys like GECKOS \citep{geckos} aim to extend the analysis of $\alpha$-bimodality beyond the Milky Way. However, the Milky Way is the only known galaxy for which such a feature has been definitively known to exist.

Ages and elemental abundances act as an invaluable counterpart for kinematic data as they are independent of the dynamical history of the galaxy after birth. The content of stellar atmospheres reflects the composition of the galactic interstellar medium (ISM) at the time of star birth \citep{Tinsley}. The analysis of the stellar distributions on the [Fe/H] - [$\alpha$/Fe] plane (Tinsley-Wallerstein diagram) serve as one of the common tools for studying galactic enrichment history. This is because $\alpha$-elements are mostly deposited into the ISM via core-collapse supernovae (SNII) explosions; and while iron-peak elements are also produced to some extent in SNII, they primarily are sourced from Type Ia supernovae (SNIa), which evolve on longer timescales. The relative contribution of one group of elements over another reflects the star formation history of the galaxy. For example, intense star formation results in a large amount of SNII explosions, rapidly enriching the interstellar medium (ISM) with both alpha and iron elements. On the [Fe/H]-[$\alpha$-Fe] plane, stars born during this period would display a near-constant [$\alpha$/Fe]. 
Star formation histories that steady decline after an early peak result in a ``$\alpha$-knee'', where [$\alpha$/Fe] transitions from being nearly constant to decreasing as a function of increased metallicity; this occurs because the enrichment from SNII is no longer as productive at incorporating alpha elements in the star forming ISM and contribution from SNIa becomes more prominent \citep{Mason23}. 
Star formation history is just one of the factors affecting the shape of the stellar distribution. Among others are the gas depletion timescale, the delay time distribution (DTD) of SNIa, initial mass function (IMF), elemental abundances and rates of the outflowing and inflowing gas, and stellar yields \citep{Andrews17, Weinberg17, Gandhi2022, Chen2023, Dubay2024}. While the stellar distribution on the [Fe/H]-[$\alpha$/Fe] plane encodes a wealth of information, its interpretation is complicated.

Various mechanisms were suggested to explain the formation of high- and low-$\alpha$ sequences in the Milky Way.
In the two-infall models, proposed in \citet{Chiappini97}, the first episode of primordial gas infall results in the formation of the high-$\alpha$ old population during a strong star formation episode from well-mixed gas. The second, delayed infall feeds the galaxy on a longer timescale, dilutes the previously enriched ISM, and fuels the building of a low-$\alpha$ population with a lower star formation rate.
To reach a better agreement with observational data \citet{Spitoni2019, Spitoni2020, Spitoni2022} suggested a longer delay of the second infall episode (4.5-5.5 Gyr). 
The hiatus in the star formation history, naturally arising in their two-infall model is responsible for the rapid decrease of [$\alpha$/Fe] due to the changing ratio of SNII to SNIa \citep{Spitoni2024}. Such temporal quenching can also produce bimodal populations with non-overlapping metallicities in both the inner disk and bulge of the Milky Way \citep{Lian2020_bulge, Haywood2018_inner}.
\citet{Lian2020I, Lian2020II} suggested a modification of the two-infall model where both accretion episodes are short and lead to starburst episodes; the second gas infall in their model is delayed by 8.2 Gyr, has a timescale of 0.7 Gyr and could be related to the gas stripping from Sgr \citep{TepperGarcia2018}.
Some models suggest that the gas accreted during the second infall might not be directly associated with merger events, but could come from the surrounding gaseous halo that was enriched by feedback-driven outflows during the early thick disk formation \citep{Khoperskov21, Haywood2019}.

Models centered around the radial migration mechanism do not require dilution of ISM 
to form the low-$\alpha$ sequence. 
While in two-infall models the evolutionary track on the [Fe/H]-[$\alpha$/Fe] plane creates a loop following the accretion of metal-poor gas, in radial migration models the evolutionary tracks show a monotonic decrease of [$\alpha$/Fe] with increasing [Fe/H]. However, the metallicity gradient in the disk ensures that stars at different radii form with different values of [Fe/H] and [$\alpha$/Fe]. Due to scattering processes, stars born at other radii can migrate to the solar neighbourhood, resulting in the formation of low-$\alpha$ sequence with a spread in metallicity. The gap between populations with distinct level of [$\alpha$/Fe] is due to the delayed onset of Type Ia supernovae.
Chemical evolution models with radial migration and radial flows of gas were explored in multiple works and have been able to reproduce various trends of [Fe/H] and [$\alpha$/Fe] with radius, vertical distance, and age; however, there is still no consensus on the strength of radial migration in the Milky Way \citep{SchonrichBinney2009, Loebman2011, Kubryk2015, Sharma2021, Johnson2021, Chen2023, Prantzos2023}.

In addition to analytical chemical evolution models, cosmological hydrodynamical simulations provide a complementary perspective on the formation of bimodal patterns and allow an estimation of how common this feature could be, however they do not really agree with the analytical models above. 
Based on the analysis of 133 galaxies from the EAGLE simulations, \citet{Mackereth2018} concluded that bimodal features are rare. Only 5 percent of galaxies from their sample showed prominent bimodality explained by distinct episodes of gas accretion. For six highest resolution galaxies in the Auriga suite, \citet{Grand18_Auriga} identified two main scenarios producing bimodality: (1) a centralised starburst, induced by an early gas-rich merger, followed by low-level star formation or (2) shrinking of the gas disk after the formation of the high-$\alpha$ disk followed by metal-poor accretion. In the VINTERGATAN simulation of one Milky Way-mass galaxy \citep{Vintergatan1, Vintergatan3}, the low-$\alpha$ sequence is formed from gas deposited by the metal-poor intergalactic filament tilted relative to the previously formed high-$\alpha$ and metal-rich disk. In four high-resolution galaxies from the NIHAO-UHD simluations, the formation of the low-$\alpha$ sequence is attributed to gas-rich merger events that bring fresh gas \citep{Buck20}.
\citet{Chandra2023} found a Milky Way analog matching the observations on the circularity-metallicity plane in the IllustrisTNG50 cosmological simulations; in their case the spatially thick disk and in-situ halo component were formed after the major gas-rich merger that tilted and heated the kinematically cold old disk. Analysis of another Milky Way analog from IllustrisTNG50 in \citet{Beane2024II} suggests that $\alpha$-bimodality could be produced without any merger event, due to the brief suppression of star formation following the formation of a bar and subsequent AGN activity.
Another independent pathway to bimodality was discovered in idealized simulations in \citet{Clarke19}. They posit a ``clumpy model'' in which the gas-rich clumps experience strong bursts of star formation and quickly self-enrich to high-$\alpha$, while the distributed mode of star formation in the disk creates a low-$\alpha$ population.
Analysis of the stellar abundances for lower-mass FIRE-2 galaxies shows that they also can exhibit bimodal features; in one of eight dwarfs examined in \citet{Patel22} a low-$\alpha$ sequence emerges due to the strong burst of star formation, followed by quenching period, during which iron enrichment from Type Ia supernovae efficiently lowers [Mg/Fe].

In this paper we analyse the chemical abundance space of the disks of 11 Milky Way-like galaxies from the FIRE-2 (Feedback In Realistic Environment) suite of simulations and demonstrate the effect of the late-time accretion of metal-poor gas on the presence of bimodal patterns; four additional galaxies with stronger bimodality are examined in detail in Barry et al. (in prep). In section \ref{sims} we describe our sample of simulations, and in section \ref{res} we characterize the abundance patterns, recent gas accretion history, and sources of accreted gas. In section \ref{disc} we discuss this analysis in the context of previous research on $\alpha$-bimodality in simulations and we conclude in section \ref{sum}.

\section{FIRE-2 simulations}\label{sims}
For analysis we use 11 Milky Way-like galaxies from the FIRE-2 cosmological baryonic zoom-in simulations \citep{Hopkins_FIRE2}. This includes 5 cosmologically isolated galaxies from the \textit{Latte} suite, presented in \citet{Wetzel_latte, refA, Hopkins_FIRE2, refD}, and 3 Local Group-like pairs from ``ELVIS on FIRE'' \citep{refD, refE}. 

A detailed description of the FIRE-2 physics model can be found in \citet{Hopkins_FIRE2}, but we summarize some important aspects here. The simulations were run with the code GIZMO \citep{gizmo}, which employs a mesh-free finite-mass Lagrangian Godunov method, combining advantages of both grid-based and particle-based approaches. The baryon particle mass resolution is 7070 $M_{\odot}$ for \textit{Latte} galaxies and approximately two times better for ELVIS simulations (4000 $M_{\odot}$ for \texttt{Romulus}/\texttt{Remus} and \texttt{Thelma}/\texttt{Louise} and 3500 $M_{\odot}$ for \texttt{Romeo}/\texttt{Juliet}).  

The simulations implement metallicity-dependent heating and cooling processes for gas particles in a temperature range from 10 to $10^{10}$ K, accounting for free-free, photoionization, recombination, Compton, photo-electric, metal-line, molecular, fine-structure, dust collisional, and cosmic ray processes. All runs also include a uniform redshift-dependent UV background \citep{FG2009}. 
Star formation occurs in Jeans unstable, self-gravitating, and self-shielding gas with density exceeding $n_{\mathrm{crit}} = 1000~\mathrm{cm}^{-3}$. When star formation criteria are met, a gas particle turns into a star particle that inherits the mass and elemental composition from the progenitor and represents a single stellar population with zero age on the main sequence. The star particle is assumed to have an IMF from \citet{KroupaIMF} and evolve according to stellar population models from STARBURST99 \citep{starburst99}. Explicit treatment of stellar feedback is one of the most important features of the FIRE simulations, and it includes explosions of supernovae of Type Ia and II, winds from OB and AGB stars, photo-ionization and photo-electic heating, and radiation pressure. 

Type Ia explosions occur with a rate-per-unit-stellar-mass from \citet{Mannucci_2006}, including prompt and delayed populations:
\begin{equation}
    \begin{array}{l}
        dN_{\mathrm{Ia}}/dt = 0\mathrm{~for~}t_{\mathrm{Myr}} < 37.53 \\ 
        dN_{\mathrm{Ia}}/dt = 5.3 \times 10^{-8} + 1.6 \times 10^{-5} \exp \{-[(t_{\mathrm{Myr}} - 50) / 10]^2\} \\ \mathrm{~for~}t_{\mathrm{Myr}} \geq 37.53 \\ 
    \end{array}
\end{equation}
Yields for SNe Ia are from \citet{Iwamoto}.
Rates-per-unit-stellar-mass of SN II explosions can be approximated with the following functions:
\begin{equation}
    \begin{array}{l}
        dN_{\mathrm{II}}/dt = 0\mathrm{~for~}t_{\mathrm{Myr}} < 3.401\mathrm{~or~} t_{\mathrm{Myr}} > 37.53\\ 
        dN_{\mathrm{II}}/dt = 5.408 \times 10^{-4} \mathrm{~for~}3.4 < t_{\mathrm{Myr}} < 10.37 \\ 
        dN_{\mathrm{II}}/dt = 2.516 \times 10^{-4} \mathrm{~for~}10.37 < t_{\mathrm{Myr}} < 37.53 \\ 
    \end{array}
\end{equation}
Yields for SN II follow the tables from \citet{Nomoto}. The mass and time resolution of the simulations are sufficient to resolve individual supernovae events rather than approximate their collective effect. 
Metals, momentum, mass, and energy from stellar mass-loss are deposited into surrounding gas particles at every timestep and depend on the metallicity and age of the star particle. Yields for stellar winds are from the models described in \citet{vandenHoek, Marigo2001, Izzard},  and are synthesized in \citet{Wiersma}. The simulations trace 11 elements (H, He, C, N, O, Ne, Mg, Si, S, Ca, Fe), which is crucial for the present analysis of chemical evolution; about half of the iron in this model is produced by SNe II and about half by SNe Ia.  

Additional physics include metal diffusion to account for the enrichment from neighbouring gas resolution elements, which is implemented with the following model examined in \citet{Shen2010}:
\begin{equation}
\begin{array}{c}
\displaystyle\frac{\partial \mathrm{M}_i}{\partial t}+\nabla \cdot\left(D \nabla \mathrm{M}_i\right)=0,\\ 
    D=C_0\|\mathrm{\mathbf{S}}\|{ }_f \mathrm{\mathbf{h}}^2, 
\end{array}
\end{equation}
where $\mathrm{\mathbf{h}}$ is the resolution scale, $\mathrm{\mathbf{S}}$ is the shear tensor, and $C_0$ is a constant calibrated from numerical simulations \citep{Su2017}. The details of the implementation are described in \citet{Hopkins_FIRE2}; the impact of metal diffusion on enhancing the agreement between simulated and observed metallicity distributions in the low-mass galaxies is examined in \citet{Escala2018}.

All galaxies experience disk settling around $z \sim 1$ \citep{Ma2017, GK2018, McCluskey2024}. Early star formation has a bursty nature, which later transitions to a steady mode \citep{Yu21, Gurvich23, Parul23}. Table \ref{tab:radii} lists present-day stellar masses, halo masses, and sizes, along with times when bursty star formation ends.

\begin{table*}
    \centering
    \begin{tabular}{c|c|c|c|c|c|c|c|c|c|c}
     Galaxy & $M_{\ast}$, $M_{\odot}$ & $M_{\mathrm{halo}}$, $M_{\odot}$ &  $m_i$, $M_{\odot}$ & $t_{\mathrm{bursty}}$, Gyr & $t_{\mathrm{low-\alpha}}$, Gyr & $R_{50,\ast}$, kpc & $R_{98,\ast}$, kpc & $R_{\mathrm{edge},\mathrm{gas}}$, kpc &  $R_{\mathrm{infall}}$, kpc & Reference  \\
     \hline
     \texttt{m12i} & $6.4\times10^{10}$ & $1.2 \times 10^{12}$ & 7070 & 3.14 & - & 2.6 & 15.8 & 18.8 & 25 & B \\
     \texttt{m12m} & $1.1\times10^{11}$ & $1.6 \times 10^{12}$ & 7070 & 3.81 & - & 4.3 & 19.8 & 18.8 & 30 & C \\
     \texttt{m12c} & $6.0\times10^{10}$ & $1.4 \times 10^{12}$ & 7070 & 3.70 & - & 2.9 & 15.8 & 16.4 & 25 & D \\
     \texttt{m12f} & $8.6\times10^{10}$ & $1.7 \times 10^{12}$ & 7070 & 5.01 & - & 2.9 & 24.8 & 27.8 & 35 & A\\
     \texttt{Thelma} & $7.9\times10^{10}$ & $1.4 \times 10^{12}$ & 4000 & 2.57 & - & 3.4 & 19.9 & 25.5 & 30 & D \\
     \texttt{Louise} & $2.9\times10^{10}$ & $1.1 \times 10^{12}$ & 4000 & 5.56 & 3.37 & 2.8 & 23.2 & 38.8 & 45 & D \\
     \texttt{Juliet} & $4.2\times10^{10}$ & $1.1 \times 10^{12}$ & 3500 & 4.40 & 2.57 & 1.8 & 17.3 & 26.5 & 35 & D \\
     \texttt{Romeo} & $7.4\times10^{10}$ & $1.3 \times 10^{12}$ & 3500 & 6.52 & 3.43 & 3.6 & 24.6 & 32.9 & 45 & D \\
     \texttt{Remus} & $5.1\times10^{10}$ & $1.2 \times 10^{12}$ & 4000 & 5.88 & 3.37 & 2.9 & 23.7 & 28.6 & 35 & E \\
     \texttt{Romulus} & $1.0\times10^{11}$ & $2.1 \times 10^{12}$ & 4000 & 4.90 & 4.19 & 3.2 & 26.6 & 32.1 & 35 & E \\
     \texttt{m12b} & $8.1\times10^{10}$ & $1.4 \times 10^{12}$ & 7070 & 6.32 & 5.80 & 2.2 & 18.7 & 18.5 & 35 & D \\
    \end{tabular}
    \caption{Properties of galaxies from FIRE-2 simulations used in this paper. We provide the stellar mass ($M_{\star}$) within the central 20 kpc at z = 0, the halo virial mass ($M_{\mathrm{halo}}$) defined by $\rho_{200\mathrm{m}}$, the mass resolution for baryonic particles ($m_i$), the lookback time to the transition from bursty to steady star formation ($t_{\mathrm{bursty}}$) derived in \protect\citet{Yu21}, the radius enclosing 50\% of stellar mass at z = 0 ($R_{50,\ast}$), the radius enclosing 98\% of stellar mass at z = 0 ($R_{98,\ast}$), the radius where the surface density of gas falls below 5 $M_{\odot}/\mathrm{kpc}^2$ ($R_{\mathrm{edge},\mathrm{gas}}$), the radius of the 10-kpc shell enclosing gas infall particles ($R_{\mathrm{infall}}$).\\\hspace{\textwidth}
    The references are: A: \protect\citet{refA}, B: \protect\citet{Wetzel_latte}, C: \protect\citet{Hopkins_FIRE2}, D: \protect\citet{refD}, E: \protect\citet{refE}}
    \label{tab:radii}
\end{table*}

\section{Results}\label{res}

\subsection{Abundance patterns in the sample}\label{res:patterns}

We begin with an examination of the stellar distribution on the [Fe/H]-[$\alpha$/Fe] plane and classification of the simulated galaxies based on their abundance patterns. For the analysis, we used disk stars that formed inside the physical radius enclosing 98\% of stars at present day, $R_{\ast, ~98}$, which is listed in the Table~\ref{tab:radii}. Disk stars were selected based on the circularity, $\varepsilon = j_z/j_{\mathrm{circ}} > 0.2$ following the criteria in \citet{Yu21}, where $j_{\mathrm{circ}}$ is the angular momentum of the circular orbit with the same total energy \citep{Abadi03}. We experimented with various sizes of the radial cut and found that the conclusion about the presence or lack of bimodality remains unchanged for smaller aperture size. However, including stars in the outermost parts helps to reveal more clearly the metal-poor tail of the low-$\alpha$ sequence in bimodal galaxies. One concern is that the large aperture size might lead to a non-negligible fraction of accreted stars, but those stars should create separate tracks on the [Fe/H]-[$\alpha$/Fe] plane that do not align with high- and low-$\alpha$ sequences.

\begin{figure*}
    \includegraphics[width=\textwidth]{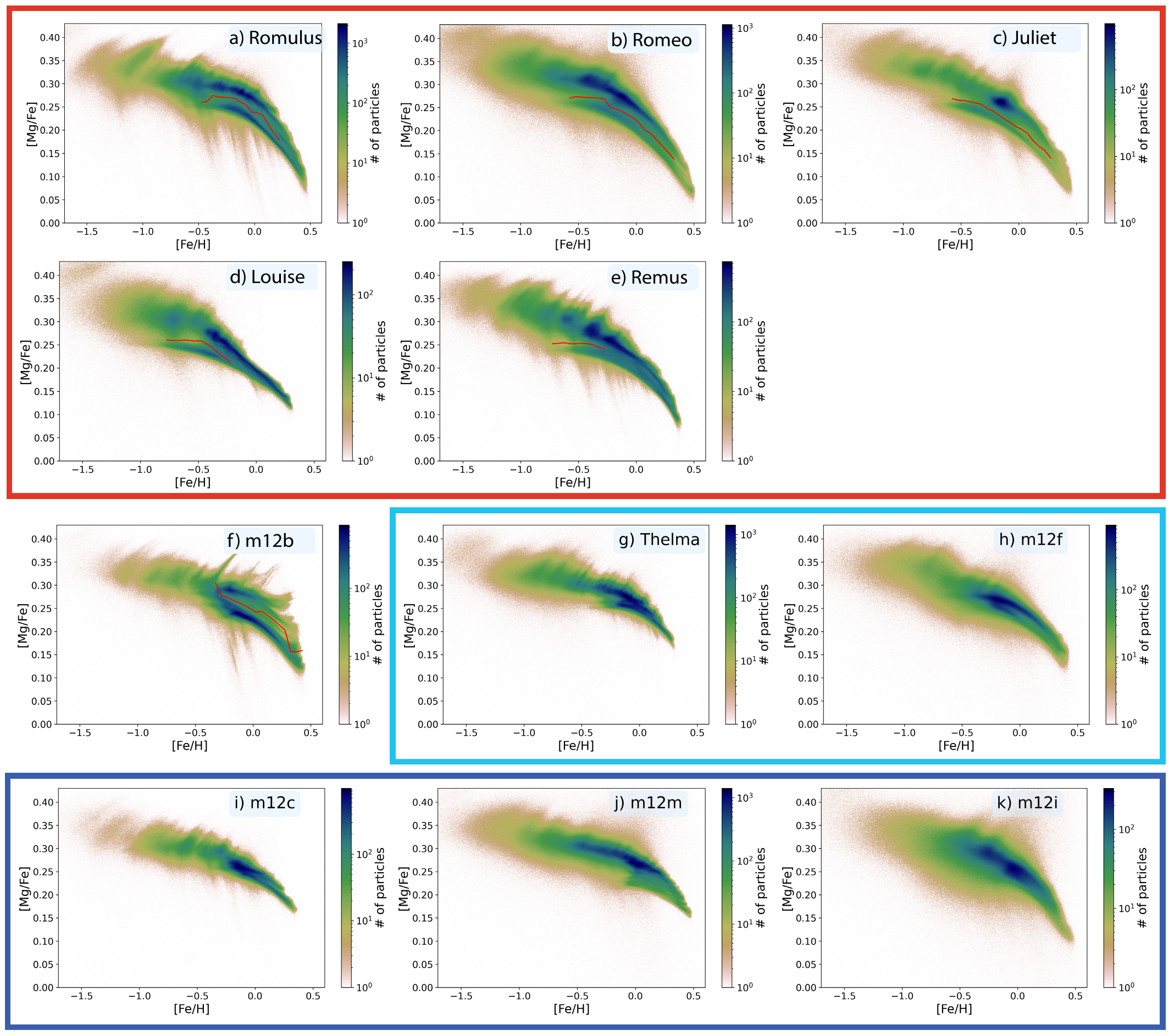}
    \caption{Distributions on the [Fe/H]-[Mg/Fe] plane for disk stars with formation radius smaller than $R_{98}$ at $z=0$. Panels (a) - (e) in red frame show bimodal galaxies, (f) -- ``extreme'' case with multiple sequences, (g), (h) in blue frame -- weakly bimodal galaxies, (i) - (k) inside navy frame -- non-bimodal galaxies. The red curve on panels (a) - (e) shows the separation between high- and low-$\alpha$ sequences.}
    \label{fig:feh_mgfe}
\end{figure*}

In Fig.~\ref{fig:feh_mgfe} we plot the distribution of the disk stars born within $R_{98}$ on the [Fe/H]-[$\alpha$/Fe] plane for 11 Milky Way-like galaxies, with Mg as a representative $\alpha$-element, because it is the most ``purely'' created by CCSN in the FIRE-2 model. 
Galaxies display a variety of complex patterns, that resemble the $\alpha$-bimodality in the Milky Way, i.e. the presence of two distinct sequences with different levels of [Mg/Fe]. The red curve on panels (a) -- (e) of Fig.~\ref{fig:feh_mgfe} separates the whole sample into high and low-$\alpha$ sequences. We describe the construction of this threshold below.

We select stars in narrow ($\Delta\mathrm{[Fe/H]} = 0.05$) bins by [Fe/H] and compute the distribution of [Mg/Fe] in each bin. We then use the \texttt{scipy.signal.find\_peaks} function to locate the peaks in the normalized distribution with a prominence of at least 0.3 and a distance between peaks of at least 0.025 dex, which is set by the typical uncertainty of abundance measurements in the Milky Way \citep{Ness2018, Buder2021}. 
If two peaks are detected, the [Mg/Fe] value at the minimum between two peaks is taken as the threshold value between the two sequences in the current [Fe/H] bin.

\begin{figure*}
    \centering
    \includegraphics[width=0.5\textwidth]{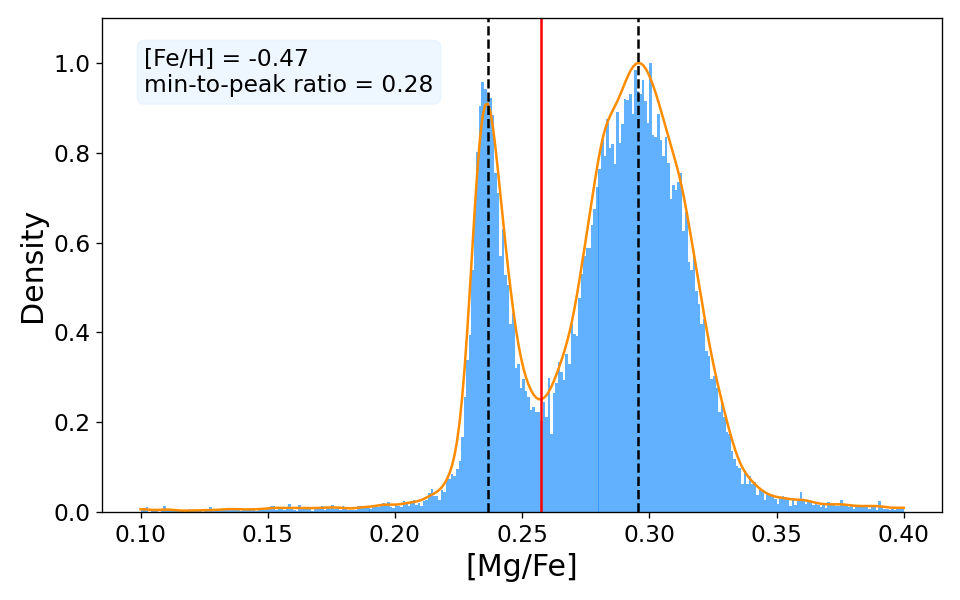}
    \hfill
    \includegraphics[width=0.45\textwidth]{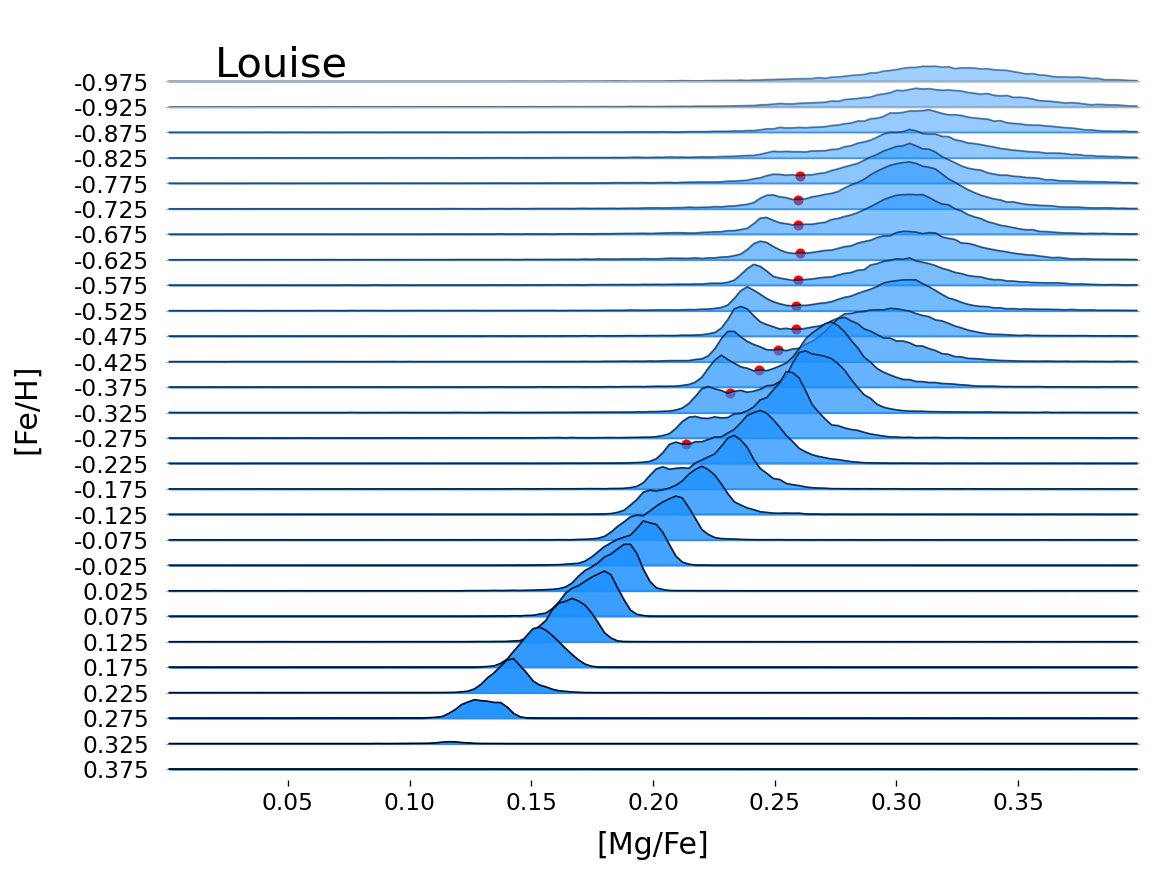}
    \caption{\textit{Left:} distribution of $\mathrm{[Mg/Fe]}$ for stars with $\mathrm{[Fe/H]} =-0.47 \pm 0.05$ in \texttt{Louise}. Black dashed lines indicate the location of the peaks in the distribution, red lines marks the location of the minimum between two peaks, separating stars belonging to the low- and high-$\alpha$ sequences. \textit{Right:} distribution of $\mathrm{[Mg/Fe]}$ for stars binned by $\mathrm{[Fe/H]}$, red dots mark the position of minimum, when two peaks are detected. \texttt{Louise} has merging sequences that are separated at $\mathrm{[Fe/H]} \lesssim -0.32$}
    \label{fig:peaks}
\end{figure*}

Figure ~\ref{fig:peaks} illustrates the process: the left panel shows the distribution of [Mg/Fe] for stars with $\mathrm{[Fe/H]}=-0.47 \pm 0.025$, while the location of the peaks and minimum between them are marked by black dashed and red solid lines, respectively. The right panel shows the distribution of [Mg/Fe] across all bins of [Fe/H] with red dots marking the location of the minima when they are detected in the bin; Fig.~\ref{fig:all_joy} demonstrates the distributions for the rest of the galaxies in the sample.

We label galaxies \texttt{Louise}, \texttt{Remus}, \texttt{Romulus}, \texttt{Romeo}, and \texttt{Juliet} as ``bimodal'', since they have double peaks in their [Mg/Fe] distribution across several [Fe/H] bins. In \texttt{Romulus}, \texttt{Romeo}, and \texttt{Juliet} two sequences are almost parallel to each other (``genuinely bimodal'' subcategory), while in \texttt{Louise} and \texttt{Remus} two sequences are merging at [Fe/H] $\gtrsim -0.25$ (``converging bimodal'' subcategory), but in all five galaxies bimodality is clear in low-metallicity regime.
Galaxies \texttt{Thelma and m12f} show very subtle bimodal features, with double peaks detected only in a narrow range of metallicities when the criteria for distance between peaks is lowered from 0.025 to 0.015. We label them  as ``weakly-bimodal''. Galaxies \texttt{m12m, m12i, m12c} are examples of ``non-bimodal'' case.

One unusual example is galaxy \texttt{m12b} which has three peaks at solar and super-solar metallicities; also the low-$\alpha$ sequence in \texttt{m12b} doesn't reach as low [Fe/H] as in our bimodal galaxies. 
We label this galaxy as ``strongly''-bimodal and defer the discussion of this case and other strongly-bimodal galaxies (not presented here) to Barry et al (in prep).

For the bimodal galaxies we define the onset of low-$\alpha$ sequence as a lookback time, when fraction of the stars forming in the low-$\alpha$ sequence (i.e. below the red line on Fig.~\ref{fig:feh_mgfe}) exceeds 20\%, since after reaching this threshold the fraction rises steeply as could be seen on Fig.~\ref{fig:fraction_low}. To select low-$\alpha$ stars in ``converging bimodal'' galaxies (\texttt{Louise} and \texttt{Remus}) we extrapolated the separation line (red line on Fig.~\ref{fig:feh_mgfe}) into the converging region.
The onset times are listed in Table~\ref{tab:radii}.

\subsection{Temporal and spatial trends of abundance patterns}

\begin{figure*}
    \centering
    \includegraphics[width=.48\linewidth]{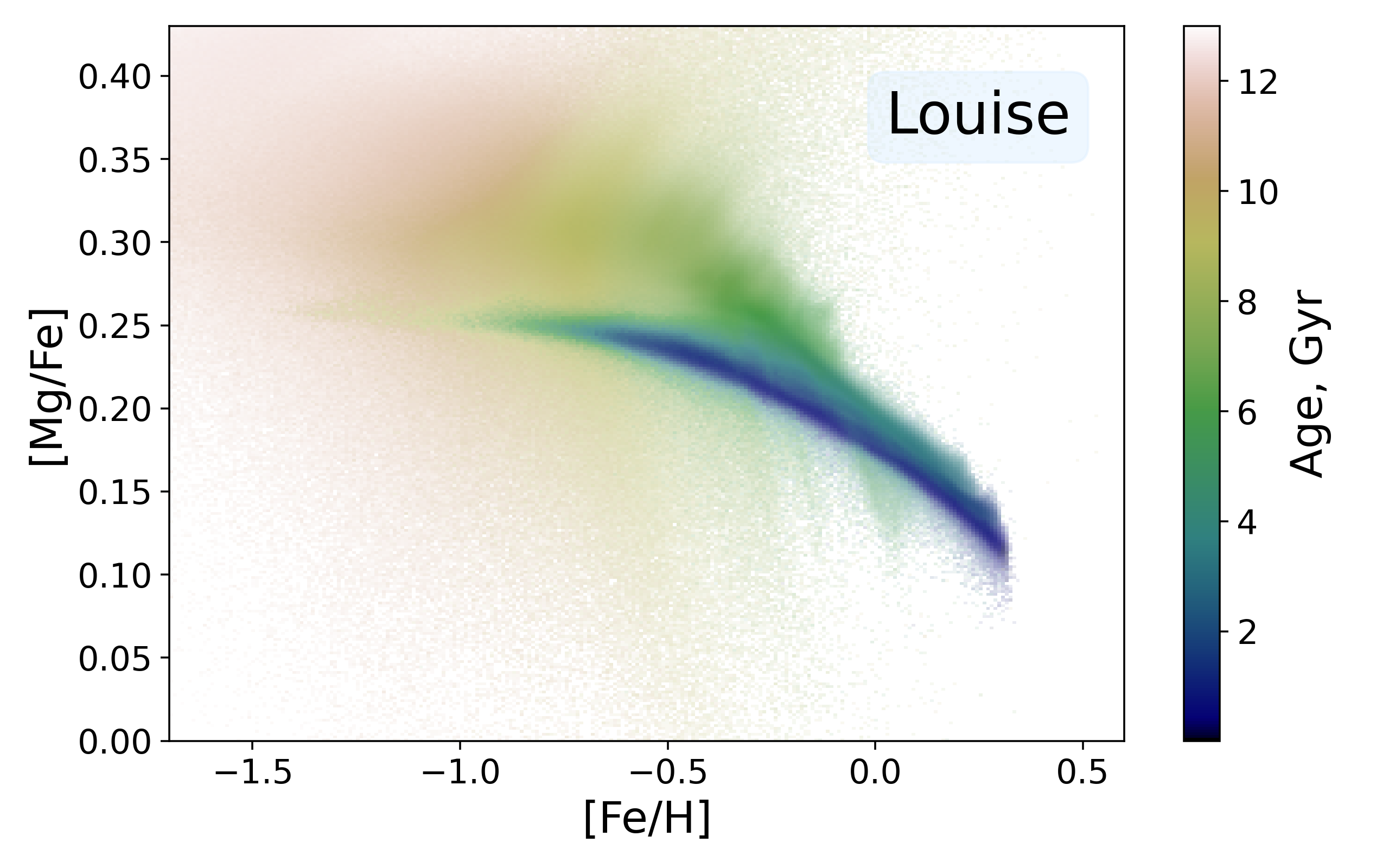}
    \includegraphics[width=.48\linewidth]{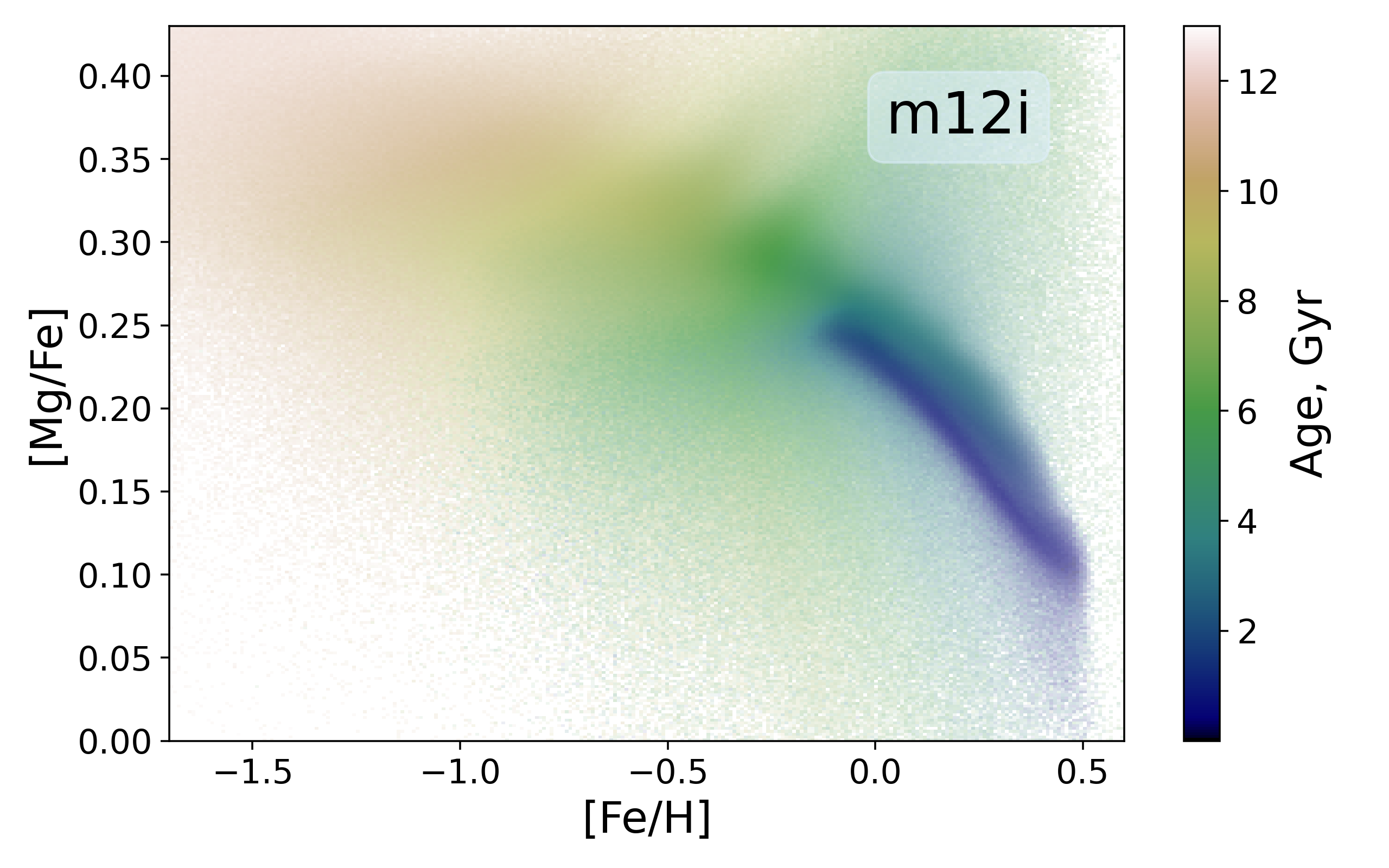}
    \caption{2D histograms of distribution on [Fe/H] - [Mg/Fe] colored by age of stars for \texttt{Louise} (left) and \texttt{m12i} (right). In both galaxies the high-$\alpha$ sequence has a wide scatter in [Mg/Fe] and is formed by old stars, and then becomes narrower for stars younger than $\approx$6 Gyrs.} 
    \label{fig:Louise_m12i_chem_age}
\end{figure*}

In this section, we describe the temporal and spatial trends on the [Fe/H]-[Mg/Fe] plane. To compare the trends between galaxies from different categories we selected \texttt{Louise} as an example of bimodal galaxy and \texttt{m12i} as a non-bimodal case.  
Fig.~\ref{fig:Louise_m12i_chem_age} shows the stellar distributions for the in-situ stars in \texttt{Louise} (left) and \texttt{m12i} (right). Distributions are color-coded by age of the stars, with transparency representing the number density of star particles in a given bin. All galaxies, regardless of the presence of bimodal features, evolve on the abundance plane in a
similar manner. Stars older than 5-7 Gyrs are located on the plateau of the high-$\alpha$ sequence, which has a wide scatter in [Mg/Fe] with prominent diagonal stripes linked to the episodes of star formation bursts in the early epoch \citep{Parul23, Patel22}. Near-constant (in \texttt{Louise} at 10-8 Gyr) or slowly decreasing (in \texttt{m12i}) level of [Mg/Fe] results from high star formation rate, when enrichment mainly comes from SN II with yields dominated by $\alpha$-elements rather than iron. At later epochs, characterized by more moderate star formation, the contribution from SN Ia explosions increases, consequently lowering the relative abundance of magnesium, [Mg/Fe], due to iron-rich yields. The low-$\alpha$ sequence in \texttt{Louise} is exclusively composed of young stars spanning a wide [Fe/H] range, indicating the presence of a radial metallicity gradient in the disk.

\begin{figure*}
    \centering
    \includegraphics[width=.48\linewidth]{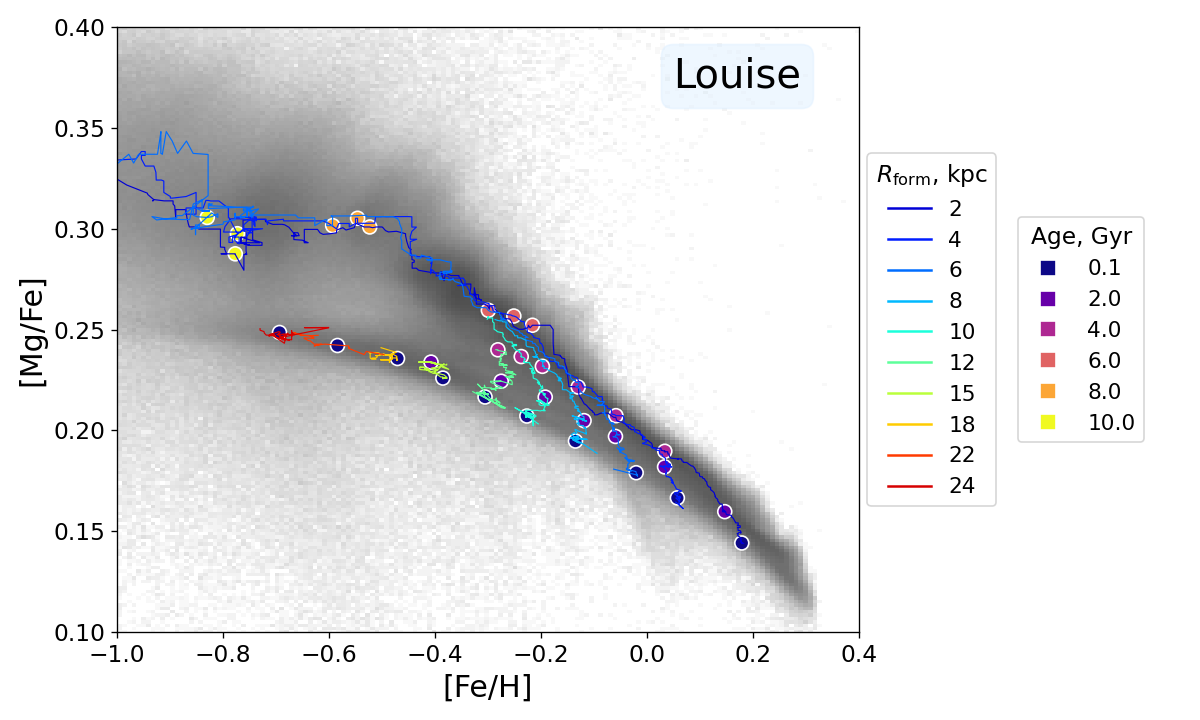}
    \includegraphics[width=.48\linewidth]{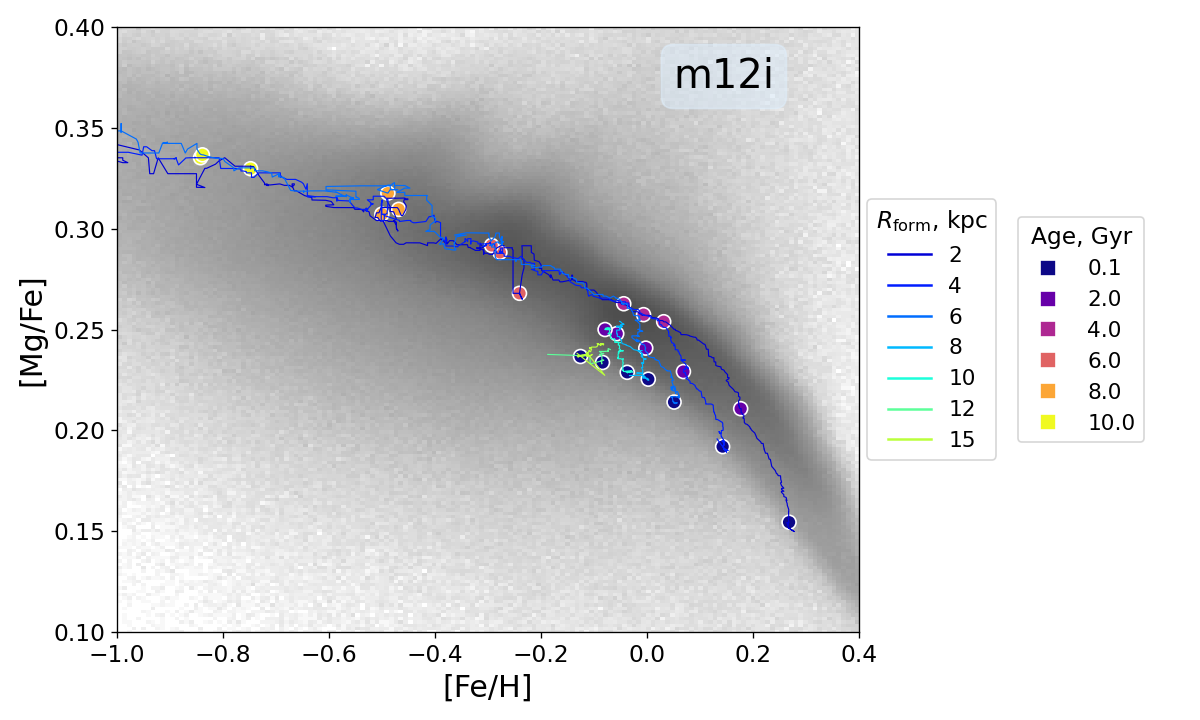}
    \caption{Elemental evolution tracks for stars binned by birth radii in \texttt{Louise} (left) and \texttt{m12i} (right). The dots mark the age of the stars with corresponding median [Fe/H] and [Mg/Fe]. The greyscale histogram in the background shows the total distribution of the disk stars formed within $R_{\ast,~98}(z=0)$. The early evolution is characterised by rapid metal enrichment with nearly constant or only slightly decreasing [Mg/Fe]. At later times, both galaxies develop a radial gradient in metallicity and at each radial bin evolve towards smaller [Mg/Fe]. While the inner radii show enrichment in [Fe/H], outer radii evolve at almost fixed metallicity. Bimodality in \texttt{Louise} is visible mainly at outer radii.}
    \label{fig:Louise_m12i_track}
\end{figure*}

To consider simultaneously the age and spatial structure of the [Fe/H]-[Mg/Fe] distributions, we plotted the evolutionary tracks of the median [Fe/H] and [Mg/Fe] for stars born within 0.1 kpc-wide radial bins on Fig.~\ref{fig:Louise_m12i_track} for \texttt{Louise} (left) and \texttt{m12i} (right).
To account for the inside-out growth of the disk, we first measure the time evolution of $R_{\ast,~98}$ -- the radius enclosing 98\% of stars. Then, for each radial bin $R$, we define the time interval when $R_{\ast,~98} > R$ and plot only that portion of the evolutionary track. The dots on Fig.~\ref{fig:Louise_m12i_track} mark the age of the stars born at specific radius with specific median [Fe/H] and [Mg/Fe]. The tracks are overplotted on the distribution for all stars with formation radius smaller than $R_{\ast,~98}$ at z=0 (as in Fig.~\ref{fig:feh_mgfe} and Fig.~\ref{fig:Louise_m12i_chem_age}).

Since we plot the tracks only for radii smaller than $R_{\ast,~98}(t)$, Fig.~\ref{fig:Louise_m12i_track} illustrates the inside-out growth of the disk, especially striking in \texttt{Louise}, where bimodality is visible mainly at the radii, populated by young metal-poor stars ($R > 12$ kpc, [Fe/H] $< -0.3$).
Another notable feature of galaxy evolution over the last 6-7 Gyrs is the development of negative radial gradient in metallicity, which occurs in all galaxies in our sample regardless their bimodality \citep{Bellardini21, Bellardini22}, therefore formation of radial gradient alone does not define whether galaxy has bimodal patterns.

\subsection{Gas accretion history}\label{gas_accretion}
The slope of tracks of median [Mg/Fe]-[Fe/H] steepens at the most recent 5-6 Gyr, reflecting the rapidly changing ratio of Type Ia to Type II supernovae following the sudden decrease of star formation rate. 
The connection between the formation of bimodal patterns and star formation history is examined in detail in Barry et al. (in prep) for the case of galaxies with strong bimodality.
However, the change in star formation is not enough to explain the behaviour of the tracks in our sample: iron enrichment from SNIa should result in evolution diagonally downward and to the right, however, some of them (e.g. \texttt{Romulus} for $R = 4$ -- $10$ kpc at age $< 4$ Gyr or \texttt{Romeo} for $R = 8$ -- $15$ kpc) drop almost perpendicularly or even shift to lower [Fe/H] (e.g. \texttt{Louise} for $R = 10$ -- $12$ kpc at age $< 2$ Gyr), meaning that iron must be either removed or diluted. Also the youngest parts of the evolutionary tracks in bimodal galaxies cover the larger range of metallicities than non-bimodal galaxies, extending far into metal-poor region. This implies that over the most recent 5 -- 6 Gyr, bimodal galaxies contained a larger reservoir of star-forming metal-poor gas, which could be a consequence of different accretion histories.

We examine the accretion in the immediate vicinity of the galaxy and define \textit{gas inflow} at a given snapshot as gas particles with negative radial velocity 
located within a 10 kpc shell around the host. The inner radius of the shell was chosen as to slightly exceed the size of the gaseous disk at the present day and is listed in Table~\ref{tab:radii}. For each galaxy, we select 15 -- 20 snapshots covering the final $\sim 8$ Gyr, which captures the entire timeframe of the low-$\alpha$ sequence evolution. Since gas cells could leave and re-enter the shell, we track them between snapshots and assign the last snapshot when the particle was inside the shell as its ``time of infall''. The actual time for a gas particle to become bound to the disk of the host galaxy could occur between our selected snapshots, later than the ``time of infall''. Therefore, the uncertainty in the accretion time is on the order of the time spacing between snapshots and we can study trends in gas accretion on timescales of approximately 0.5 Gyr. We should also note that some of the tracked gas particles could end up outside the host galaxy at the present day. However, the fraction of such particles is very small, less than 1\%. 

\begin{figure}
    \centering
    \includegraphics[width=\linewidth]{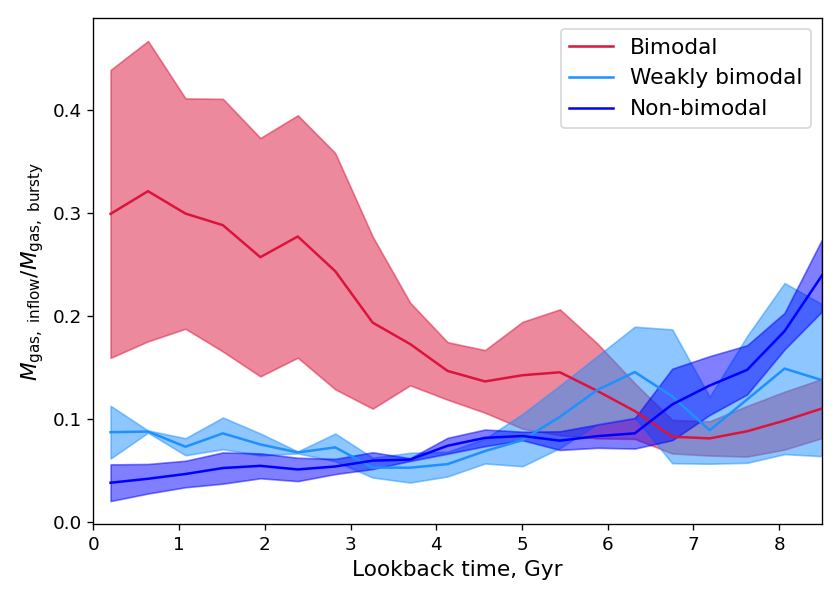}
    \caption{Time evolution of the mass of the gas inflow (see \ref{gas_accretion} for definition) normalized over the mass of gas in the galaxy at $t_{\mathrm{bursty}}$ which roughly corresponds to the disk settling time. Bimodal galaxies experience increasing inflow over the past 5-6 Gyrs.}
    \label{fig:inflow}
\end{figure}

Fig.~\ref{fig:inflow} shows the evolution of gas inflow mass; since different galaxies have different masses and we want to compare the variation in the amount of accreted gas, rather than absolute values, we have normalized the mass of the inflow by the gas mass in the host's disk measured at the end of the bursty phase of star formation, $M_{\mathrm{gas},~t_{\mathrm{bursty}}}$, when galaxies finish the formation of thick disk \citep{Yu21}.
We plot the median track for each category and show the 1$\sigma$ deviation in a shaded region.
Galaxies in bimodal and non/weakly-bimodal categories display strikingly different trends in the evolution of inflow mass. While in non-bimodal galaxies the total mass of inflowing gas is moderately declining with time, bimodal galaxies experience a rapidly increasing amount of accretion in the last $\approx$ 5 -- 6 Gyrs.

\begin{figure}
    \centering
    \includegraphics[width=\linewidth]{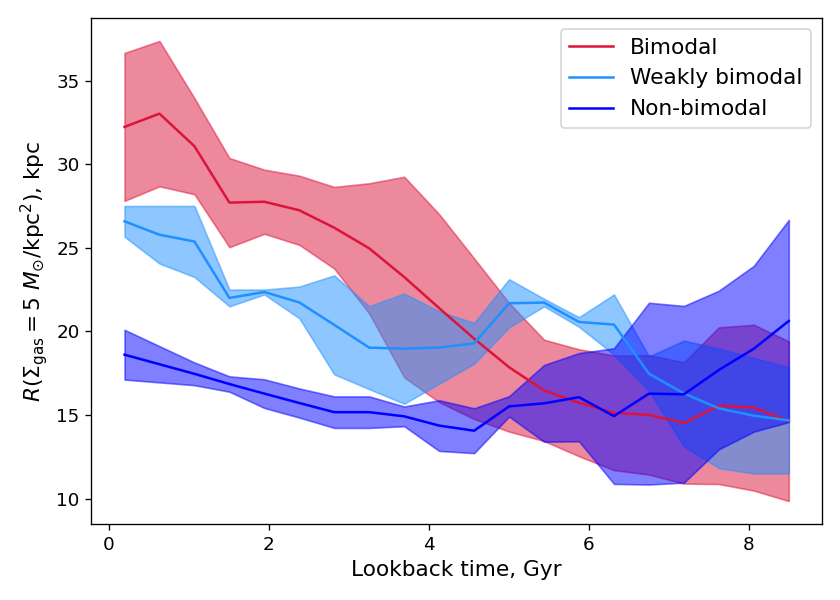}
    \caption{Time evolution of the gas disk size. Galaxies with stronger bimodal features also have more significant growth of the gas disk.}
    \label{fig:rsigma}
\end{figure}

\begin{figure}
    \centering
    \includegraphics[width=\linewidth]{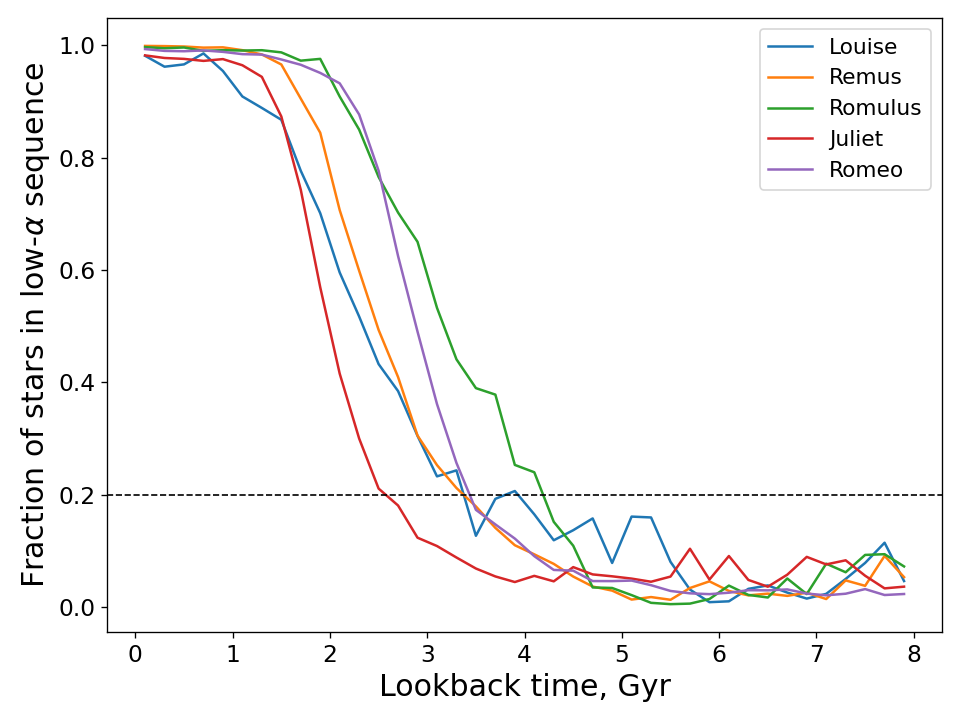}
    \caption{Fraction of stars in the low-$\alpha$ sequence for bimodal galaxies.}
    \label{fig:fraction_low}
\end{figure}

The size of the gaseous disk evolves in a similar manner. Fig.~\ref{fig:rsigma} shows the evolution of the disk edge size, which we define as the radius of the ring where the surface density falls below 5$M_{\odot}/\mathrm{kpc}^2$. At early times, the sizes of the gas disk for galaxies from different categories overlap and span a wide range of radii.
Around $\approx 5$ -- 6 Gyr ago the evolution of bimodal and non-bimodal galaxies bifurcates: bimodal galaxies gradually increase their gas disk size up to 30 -- 35 kpc, while the radius of the non-bimodal galaxies converges to $\approx$ 20 kpc, with weakly-bimodal galaxies covering the range in between. 
The fractional growth of the low-$\alpha$ populations follows the evolution of both inflow mass and gas disk size, lagging by 1 -- 2 Gyr, as shown in Fig.~\ref{fig:fraction_low}.

\begin{figure}
    \centering
    \includegraphics[width=\linewidth]{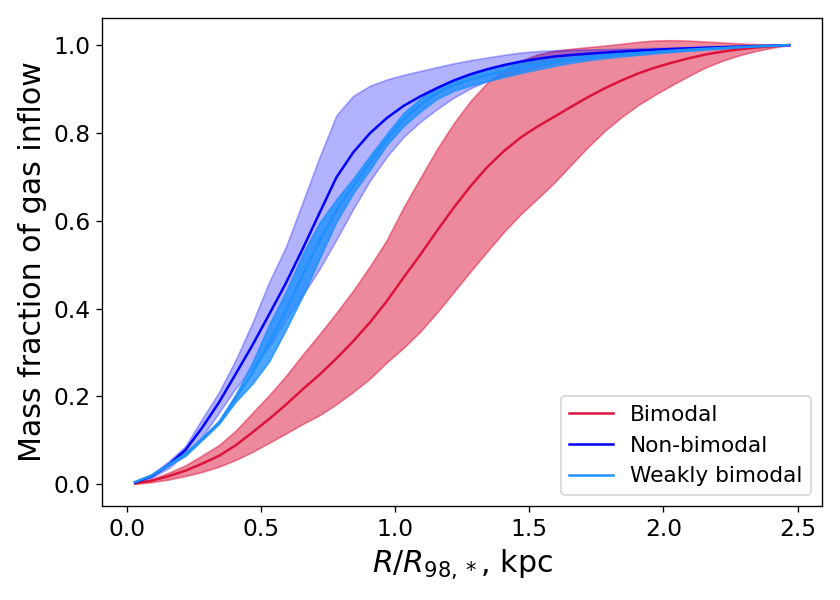}
    \caption{Radial distribution of accreted gas inflow at $z=0$, with radius normalized by $R_{98, \ast}$, the radius enclosing 98\% of the stellar mass. In bimodal galaxies, accreted inflow demonstrates a more extended spatial configuration, while in non-bimodal and weakly-bimodal galaxies most of the accreted gas is confined within $R_{98, \ast}$.}
    \label{fig:rad_distr}
\end{figure}

\subsection{Metallicity distribution of the accreted gas}
In this section we take a closer look at the properties of gas infall and link it to the observed stellar distributions on the [Fe/H]-[Mg/Fe] plane. 

\begin{figure*}
    \centering
    \includegraphics[width=\textwidth]{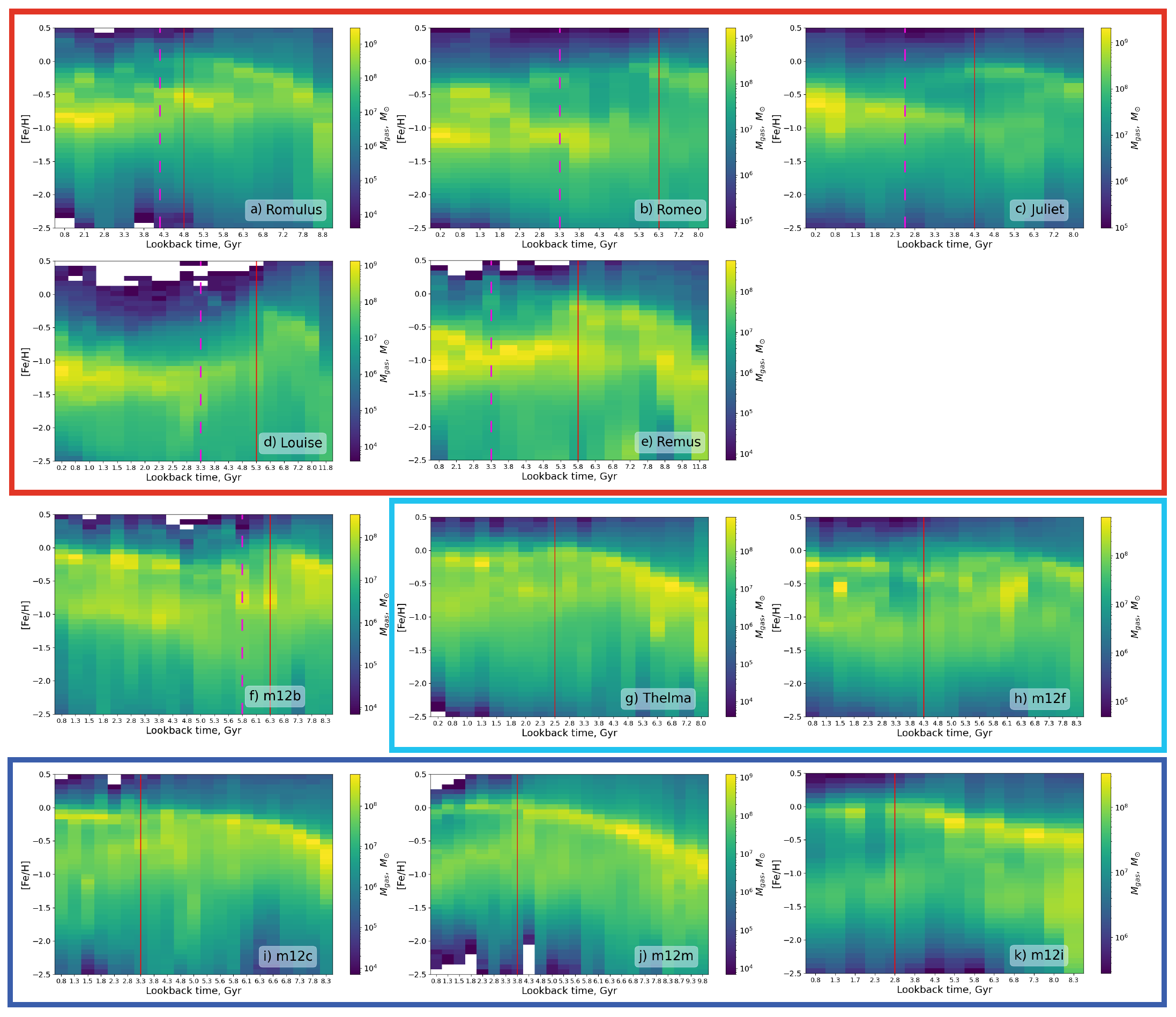}
    \caption{Mass-weighted metallicity distributions of the inflowing gas over time. The red vertical line marks the end of bursty star formation, while the pink dashed line indicates the onset of the low-$\alpha$ sequence. While all galaxies demonstrate similar distributions with a sharp cutoff at near-solar metallicities at early time, metallicity distributions of bimodal galaxies over the last 4-5 Gyrs is dominated by more metal-poor gas compared to their non- and weakly-bimodal counterparts. Red frame contains bimodal galaxies, blue frame contains weakly bimodal galaxies, and navy -- non-bimodal ones.}
    \label{fig:heatmap}
\end{figure*}

We begin with describing the metallicity of the accreted gas. Fig.~\ref{fig:heatmap} shows the metallicity distributions (MD) for the gas inflow at each snapshot for all galaxies in our sample, where the color represents the gas mass within individual metallicity bins, the pink dashed line marks the onset of the low-$\alpha$ sequence, and the red vertical line marks the end of the bursty phase of star formation in the galaxy. At early times, MDs of all galaxies in the sample look similar and display a wide and dispersed distribution with a well-defined upper boundary, strongly dominating in most of the galaxies (especially in \texttt{m12i}, \texttt{m12m}, \texttt{m12c}, \texttt{Thelma}, and \texttt{Romulus}). The locus of the upper envelope evolves from [Fe/H]$\approx$-0.6 to -0.25 and matches the median metallicity of the host galaxy at that time, which could be explained by enrichment of the inner CGM with strong outflows that are ubiquitous in the bursty phase \citep{Muratov15, Pandya2021}. In certain cases, such as \texttt{m12f} at $\approx 6$ Gyr ago, the peak of the distribution briefly shifts to lower values, which signals a gas-rich merger. At later times, the MDs of the inflowing gas in bimodal and non-bimodal galaxies become noticeably different. In non-bimodal galaxies, the upper envelope is still present, though it becomes fainter and doesn't dominate the inflow as strongly as before. For example, in \texttt{m12m} and \texttt{m12b} the distributions are bimodal over the last 3-4 Gyrs with one peak at solar metallicity (which is most likely related to the recycled winds) and another at [Fe/H]$\approx -1$. In bimodal galaxies, the upper envelope almost completely disappears and the peak of MDs abruptly shifts to much lower values. The most prominent example of this behavior is  \texttt{Louise}, where the peak metallicity decreases by $\approx 0.75$ dex on a timescale of 1 Gyr (at 5.5 Gyr ago).

To summarize, bimodal galaxies in our sample accrete a substantial amount of distinctly metal-poor gas over the final 4 -- 5 Gyr. Fig.~\ref{fig:rad_distr} shows the cumulative radial distribution of the gas inflow at $z=0$. In bimodal galaxies, accreted gas inflow demonstrates a more extended spatial configuration, which agrees with more significant growth of gaseous and stellar disks in these galaxies over the final 4 -- 5 Gyrs.
Due to the lower metallicity of the accreted gas compared to that of the host, outer radii of bimodal galaxies are populated by young metal-poor low-$\alpha$ stars (Fig.~\ref{fig:Louise_m12i_track} and \ref{fig:all_tracks} show that the elemental evolution tracks for large radii appear only at later times and reside on the low-$\alpha$ sequence). That results in the formation of the prominent metal-poor tail of the low-$\alpha$ sequence. 
Also this inflowing gas can dilute the inner radii and promote steepening of the slopes of the evolutionary tracks (most clearly seen in \texttt{Romulus}).

\subsection{Reservoir of the accreted gas}
Fig.~\ref{fig:heatmap} demonstrates that, while being metal-poor relative to the host galaxy, the accreted gas is not primordial and displays a wide range of metallicities. Here we explore the source of the inflowing gas and compare the role of mergers and smooth accretion from the CGM. 

We traced back gas inflow via particle tracking and checked whether gas particles belonged to a dark matter halo of a satellite galaxy before accretion onto the host halo. To assign gas particles to a certain dark matter halo we used cuts in position and velocity, selecting particles within 0.8$R_{\mathrm{halo}}$ with velocity within $2v_{\mathrm{circ, ~halo}}$, following \citet{Santistevan2020}.

We found that the median fraction of accreted gas that has been a part of satellite galaxies, $M_{\mathrm{sat}}/M_{\mathrm{infall}}$, calculated across all snapshots, varies from 4\% in Remus to 37\% in Juliet. The fraction is generally higher for bimodal galaxies. However, there are exceptions in both categories, such as Remus, which has $\alpha$-bimodality but shows the lowest fraction (4\%) of gas from satellites, and \texttt{m12f}, which is only weakly bimodal but has 36\% of its accreted gas of satellite origin. While in some galaxies (e.g. \texttt{Romulus}) the evolution of $M_{\mathrm{sat}}/M_{\mathrm{infall}}$ is smooth, in others (e.g. \texttt{m12f}  and \texttt{m12c}) gas with satellite origin can briefly dominate the total gas inflow and constitute up to 80\% of the inflowing mass due to a gas-rich merger or a close passage of a gas-rich satellite.

Due to relatively steady accretion of metal-poor gas from the CGM, most of the bimodal galaxies in our sample do not show signs of rapid dilution events, visible as an abrupt shift in the age-[Fe/H] relation \citep[e.g.][]{XiangRix22} or as a leftward movement of the evolutionary track on the [Fe/H]-[Mg/Fe] plane as in the scenario of the ``classic'' two-infall model \citep[e.g.][]{Spitoni2019}.
Only few radii in \texttt{Juliet} ($R = 8$ -- $12$ kpc) show some short horizontal evolution at the last 2 Gyr, following a minor merger. However, one of the weakly-bimodal galaxies, \texttt{m12f}, survives two minor mergers with total mass ratio larger than 1\% (at $\sim$6.5 Gyr ago and at $\sim$1.4 Gyr ago.), that were more gas-rich than the one in \texttt{Juliet} and left a more dramatic imprint on the elemental evolution tracks.

\begin{figure}
    \centering
    \includegraphics[width=\columnwidth]{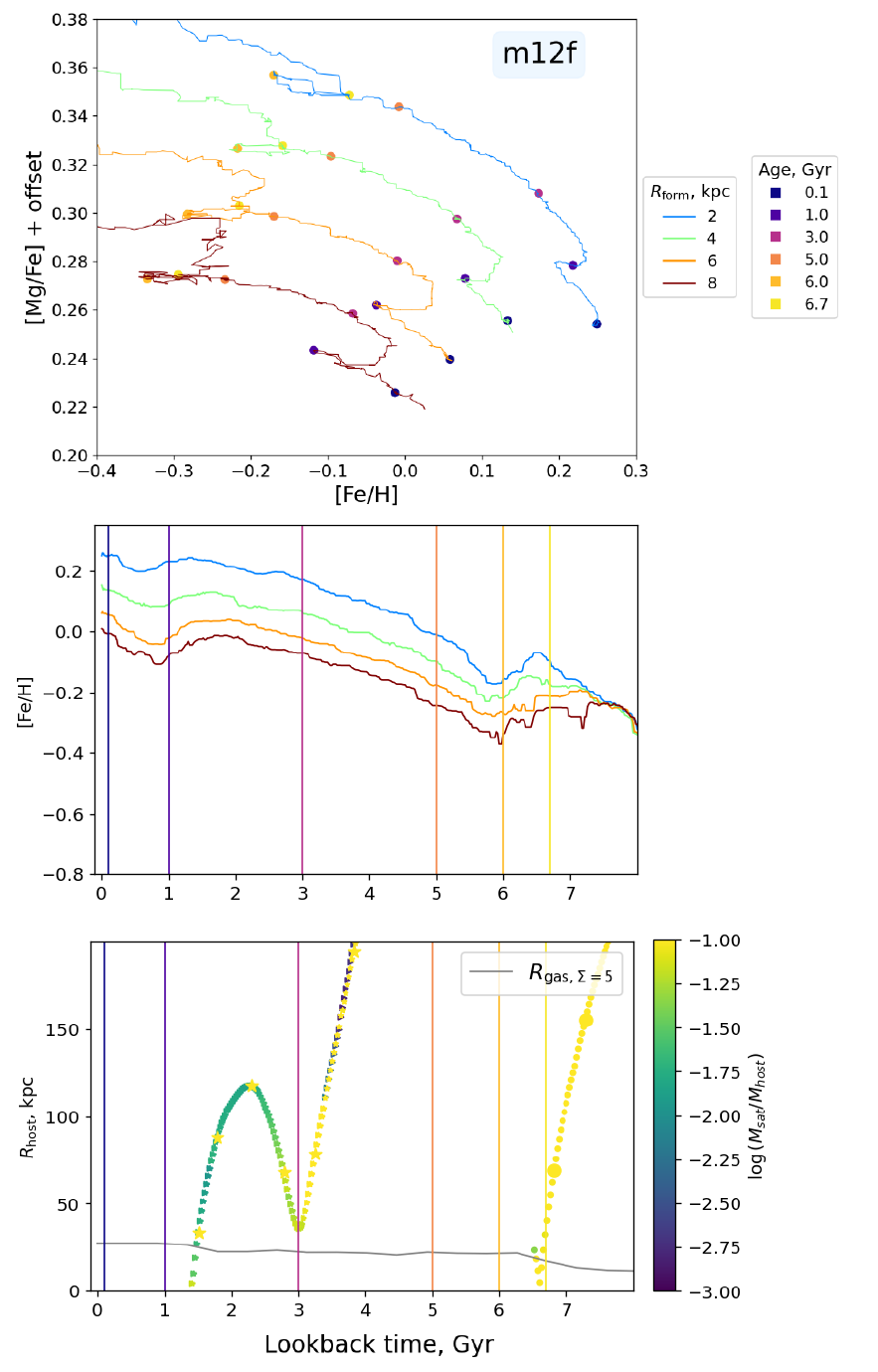}
    \caption{\textit{Top}: elemental evolution tracks for stars binned by birth radii in \texttt{m12f}. \textit{Middle}: time evolution of [Fe/H] at different radii. Colours are the same as in the top panel. \textit{Bottom}: Orbits of three satellites that contribute the most to the gas inflow, colored by mass ratio with the host galaxy. Dips in [Fe/H] at 6 and 0.75 Gyr ago follow the minor gas-rich mergers.}
    \label{fig:m12f_sat}
\end{figure}

\begin{figure*}
    \centering
    \includegraphics[width=.95\linewidth]{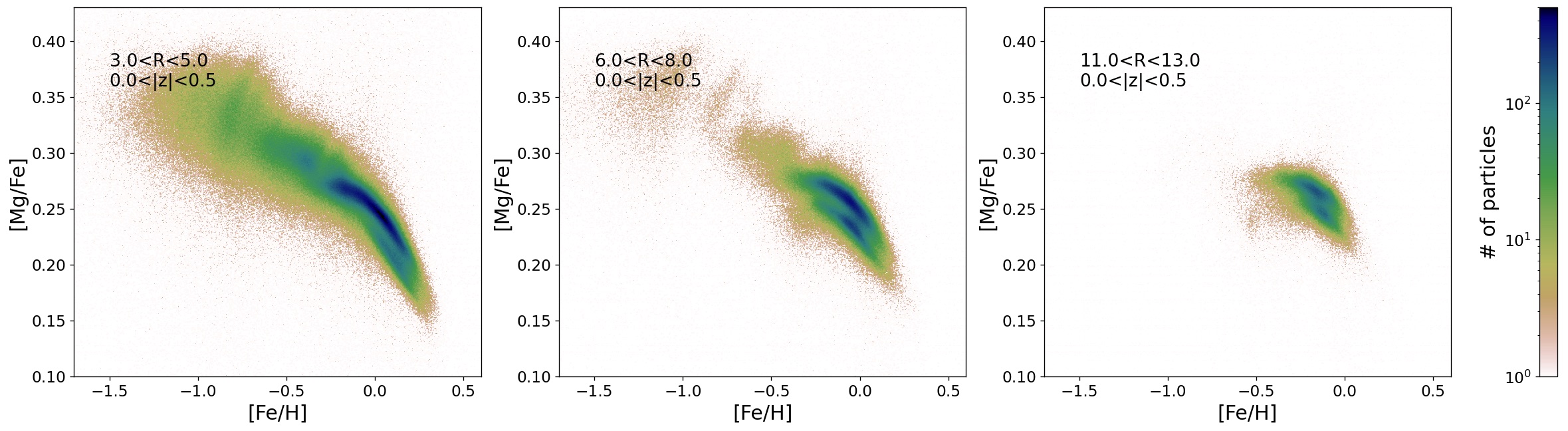}
    \caption{Distribution of stars in \texttt{m12f} on [Fe/H]-[Mg/Fe] plane in three radial bins. Despite being overall weakly bimodal, \texttt{m12f} shows very clear double sequences at certain radii.}
    \label{fig:m12f_sliced}
\end{figure*}

The top row on Fig.~\ref{fig:m12f_sat} shows the evolutionary tracks of the median [Fe/H] and [Mg/Fe] for stars born within 0.1 kpc-wide radial bins. Compared to Fig.~\ref{fig:all_tracks}, here we only show inner radii and add an offset in [Mg/Fe] to remove overlap between tracks and more clearly demonstrate the behaviour of the curves.  
The second row shows the evolution of median [Fe/H] with time at different radii with the same colour scheme as the top panel.
The bottom row shows the evolution of the distance from the host for two satellites that contribute the largest mass of gas to the inflow. The tracks are colour-coded by the relative total mass of the satellite to the host, while the colour of the markers represents the relative gas mass at the considered snapshots.

Both mergers lead to the dilution of [Fe/H], which is visible both in the [Fe/H] vs time tracks and in the horizontal evolution of the tracks on the [Fe/H]-[Mg/Fe] plane. First merger results in a narrow loop on the [Fe/H] - [Mg/Fe] plane, after which the track continues on the same path as before. After second merger, however, the track slides to the lower [Mg/Fe], creating a pattern resembling low-$\alpha$ sequence. Since this merger is very recent, the galaxy didn't have yet enough time to form a substantial amount of stars in the low-$\alpha$ sequence, so double-peaked distribution of [Mg/Fe] wasn't detected by our analysis of all in-situ stars in \ref{res:patterns}. 
However, when stars are binned by their birth radius, their distribution demonstrates a much stronger bimodal pattern, as shown on Fig.~\ref{fig:m12f_sliced}. Two parallel populations are most clearly visible for $6< R_{\mathrm{birth}}<8$ kpc, although inner and outer bins also contain two overdensities.

To explain, why the second merger results in the transition to the low-$\alpha$ sequence while the first does not, we must consider the differences in star formation history. Briefly, the second merger induced a starburst followed by a dip in star formation. In contrast, the first merger occurred during a period of bursty star formation, when the enrichment from SNIa (which should lead to a drop in [Mg/Fe]) was overshadowed by the contribution from multiple SNe II.
The detailed connection between star formation history and bimodal patterns is investigated in Barry et al. (in prep). Here, we emphasize the variety of bimodal patterns following the accretion of metal-poor gas. \textit{Smooth accretion from the CGM
primarily impacts the outer disk, resulting in a gradual change in the slope of elemental evolution tracks and a gradual build-up of the metal-poor tail of the low-$\alpha$ sequence due to star formation at large radii. More violent events, such as gas-rich mergers, can lead to rapid dilution in [Fe/H] at all radii and promote a quick transition to lower [Mg/Fe] due to changing ratio of SN Ia to SN II when the star formation history includes a quenching episode.}

\section{Discussion}\label{disc}

We explored the stellar distributions on the [Fe/H]-[Mg/Fe] plane for 11 FIRE-2 Milky Way-mass galaxies, looking for the presence of multiple sequences with distinct [Mg/Fe]. Five galaxies -- \texttt{Louise}, \texttt{Remus}, \texttt{Romulus}, \texttt{Romeo}, and \texttt{Juliet} -- show clear bimodal features, where two sequences are separated in the low-metallicity regime and either merge at solar metallicities (in the case of \texttt{Louise} and \texttt{Remus}) or run parallel to each other (as in \texttt{Romulus}, \texttt{Romeo}, and \texttt{Juliet}). Two galaxies -- \texttt{m12f} and \texttt{Thelma} -- demonstrate ``weakly bimodal'' behavior, where the range of metallicities at which two sequences are separable is relatively short. Three galaxies -- \texttt{m12i}, \texttt{m12m}, and \texttt{m12c},  -- do not have any notably distinct sequences and are classified as ``non-bimodal''. Lastly, \texttt{m12b} presents an extreme case; not only does it show multiple clearly separated tracks, but also the high-$\alpha$ sequence displays increasing [Mg/Fe] with [Fe/H] at super-solar metallicities, which is in contrast with the declining [Mg/Fe] trend seen in the rest of the galaxies (see Barry et al (in prep)).

The low-$\alpha$ sequence in our sample is built by younger stars (age $<$ 5 -- 6 Gyr) with circularities representative of stars in the thin disk. Stars in the high-$\alpha$ sequence are kinematically hotter, generally older than the low-$\alpha$ stars, and are confined to the inner regions of the galaxy, as opposed to the extended low-$\alpha$ disk. The high-$\alpha$ sequence shows a tight age-metallicity relation with younger stars being more metal-rich, resulting from the rapid enrichment at early times in a well-mixed ISM. In contrast, the age-metallicity relation in the low-$\alpha$ sequence is relatively flat and displays a wide scatter due to the metallicity gradient in the disk \citep{Bellardini22, Graf24}. 
The observed correlations between the kinematics and ages of the high- and low-$\alpha$ disks in the simulations are in qualitative agreement with trends in the Milky Way; however there are also notable differences. For example, the low-$\alpha$ disk in the Milky Way is older than in any of the simulated galaxies, and the gap in [Mg/Fe] between high- and low-$\alpha$ disks at low metallicities is larger in the Milky Way than in these FIRE-2 bimodal galaxies (a few galaxies displaying a larger gap in [Mg/Fe] are studied in Barry et al (in prep)). Additionally none of the simulated galaxies show the concave upward shape of the low-$\alpha$ sequence observed in the several Milky Way surveys \citep{Imig2023, Hayden2015, Vincenzo2021a, SilvaAguirre18, Queiroz2019}. 

To identify the conditions favorable for the formation of the low-$\alpha$ sequence, we focused on galactic evolution over the last 8 -- 9 Gyr, particularly on the differences in gas accretion histories.
We compared the mass of inflowing cold gas in the immediate vicinity of the host galaxies. While at earlier times the difference in the mass of gas inflow between bimodal, weakly bimodal, and non-bimodal galaxies is not significant, over the most recent 5 -- 6 Gyrs the bimodal galaxies experience an increasing inflow of cold gas compared to their non-bimodal counterparts. 

The metallicity of the accreted gas is not primordial and shows signatures of multiple enrichment sources that prevail at different times. At lookback times greater than 5-6 Gyrs, during the bursty phase of star formation, gas inflow is dominated by a component with [Fe/H] in the range -0.5 to 0, which is similar to the metallicity of the host galaxy at that time and likely results from the enrichment of the CGM by outflows with a high mass loading factor, ubiquitous during the bursty phase \citep{Muratov15}.
At lower redshifts, after the transition to a steady star formation mode, the contrast between the shapes of the metallicity distributions of inflowing gas in bimodal galaxies and their non-bimodal counterparts becomes more apparent. In weakly bimodal and non-bimodal systems, the metallicity distributions maintain a near-solar component, although it gets fainter with time.
In galaxies containing a low-$\alpha$ sequence, the near-solar metallicity component in the accreted gas almost completely fades, and the peak of the MDFs drops down to as low as -1.25 in cases of \texttt{Romeo} and \texttt{Louise}.

Previous works have extensively studied the process of gas accretion in FIRE-2 galaxies \citep{AnglesAlcazar_baryon}. At high redshifts, prior to halo virialization, gas accretes in a cold mode along the filamentary structure. After the inner CGM virialization, which is concurrent with disk settling and the end of bursty star formation \citep{Gurvich23}, gas from the hot halo accretes in a rotating cooling flow. As shown in \citet{Hafen22, Trapp22} the angular momentum distribution of gas accreting in the hot mode is narrow, aligned with the angular momentum of the host galaxy's disk, and slightly smaller than the value at the edge of the gas disk. We compared the difference between the specific angular momentum of the inflowing gas and the rotation curve at the edge of the disk and found that in bimodal galaxies this difference is systematically smaller, implying that the gas settles into full rotational support farther away from the galaxy center, thereby inducing a more significant growth of the disk.

Due to the accretion of a substantial amount of metal-poor gas with relatively high angular momentum, bimodal galaxies develop a large gas reservoir in their outer disks. Star formation in these outer disk regions populates the low-metallicity tail of the low-$\alpha$ sequence. \citet{Yu21} showed that thick disk formation ends with the transition from bursty to steady phase of star formation, which also coincides with the change of accretion mode. Therefore the onset of the low-$\alpha$ sequence occurs in a settled gas disk when the thick stellar disk is already in place. 
Based on the analysis of APOGEE data, \citet{Ciuca21} suggested that star formation in the outer disk occurs from the fresh infalling gas after the end of thick disk formation in the turbulent well-mixed medium, resembling the scenario realized in FIRE-2 galaxies. A similar idea of separate evolutionary paths for the outer and inner disks was proposed in \citet{Haywood2019}; they suggested that the isolation of the inner disk was provided by the outer Lindblad resonance (OLR). In our sample, the contribution of the OLR is not required -- for example, \texttt{Louise} displays the most clear bimodality with a very extended metal-poor outer disk, but doesn't form a prominent bar \citep{Ansar2023}.

Bimodal distributions resulting from a separate evolutionary path of the outer disk has been found in other cosmological simulations. In VINTERGATAN, which features one Milky Way-mass galaxy \citep{Vintergatan1, Vintergatan3}, the outer gas disk is fueled by a metal-poor cosmic filament and is initially inclined relative to the inner older thick disk. Star formation in the outer disk is induced by the passage of the last major merger, which also denotes the end of the early intense star formation activity in thick disk. Both disks later align via gravitational torquing.

Distinct episodes of gas accretion are also in the origin of $\alpha$-bimodality in the Milky Way-mass galaxies from the EAGLE simulations \citep{Mackereth2018}. The high-$\alpha$ sequence forms from gas with short consumption timescales and small radial extent. The low-$\alpha$ sequence formation proceeds over a longer timescale from gas with different enrichment evolution. They also found that the cases of bimodality are very rare and present only in 5\% of galaxies.

In the Auriga simulations \citep{Grand18_Auriga} low-metallicity gas accretion is found to be necessary to create a bimodality in the outer regions via the ``shrinking disk'' mechanism. 
In this case, the formation of high-alpha disk stars occurs early and extends out to galactocentric radii of 7 -- 12 kpc. Then for a period of roughly 1 -- 2 Gyr, star formation at large radii is halted due to gas depletion and insufficient inflow. Once the supply of external cold gas starts to increase again (mostly via minor gas-rich mergers), star formation at the outer radii resumes in the low [$\alpha$/Fe] regime and at lower metallicities than the end of the high-$\alpha$ sequence. One of the galaxies in our sample, \texttt{m12m}, also shows a shrinkage of the gas disk; however due to the lack of substantial late gas inflow, it doesn't develop a low-$\alpha$ sequence. Out of the bimodal galaxies in our sample, none shows a prominent bimodality in the outer region when stars are binned by their birth radii, hence this scenario is not significant for bimodality formation in FIRE-2.

Inner disk bimodality in Auriga galaxies arises due to a different mechanism. In the ``centralised starburst'' pathway, the high-$\alpha$ sequence forms in the compact starburst induced by early gas-rich mergers, while subsequent low level star formation creates a low-$\alpha$ sequence. Due to the enrichment from SNIa after end of the burst, gas quickly transitions to higher [Fe/H] and lower [$\alpha$/Fe] values leaving two disjoint sequences. However, inner disk bimodality doesn't show double peaks in the distribution of [$\alpha$/Fe] for any fixed [Fe/H], therefore this is a different type of bimodality than the one addressed in this paper. 

Gas-rich minor mergers are pointed out as the reason for $\alpha$-bimodality in Milky Way-mass galaxies from the NIHAO-UHD project \citep{Buck20} as they bring in fresh metal-poor gas that dilutes the ISM while keeping [$\alpha$/Fe] values mostly unchanged. 
In contrast, \citet{Khoperskov21} argues that mergers are not essential for the formation of bimodality. In their non-cosmological simulation, low-$\alpha$ sequence naturally arises due to the two-phase star formation history driven by two regimes of the gas infall. Early rapid collapse results in a strong star formation that is responsible for the building of the high-$\alpha$ thick disk, however stellar feedback efficiently removes gas from the disk and leads to the quenching of star formation. Ejected gas re-accretes on the larger timescale and participates in thin disk formation. Stars in thin disk tend to reach equilibrium values of [Fe/H] and [$\alpha$/Fe] that change with radius due to decreasing gas density at the higher radii and associated lower star formation efficiency. 

The varying nature of star formation history, rather than mergers alone, also promotes the formation of the low-$\alpha$ sequence in \citet{Beane2024}. They conducted a series of simulations with only slight variations in the orbital properties of the merger. Despite the amount and content of the deposited gas being almost identical across different runs, the simulated galaxies showed bimodal patterns only in the cases, where the star formation history included a starburst episode, followed by a quiescent phase, and subsequent rejuvenation.

One of our weakly-bimodal galaxies, \texttt{m12f}, further supports the idea that the delivery of metal-poor gas during a merger doesn't necessarily lead to a transition to the low-$\alpha$ sequence. After a second merger and its associated brief starburst, Type Ia supernovae can efficiently enrich the ISM and reduce [Mg/Fe], promoting the onset of a low-$\alpha$ sequence. However, after an earlier merger occurring during the galaxy's bursty star formation mode, the chemical evolution creates an almost horizontal loop on the [Fe/H]-[Mg/Fe] plane and then returns to the level of [Mg/Fe], making it unlikely for the galaxy to develop a bimodal structure in this regime.
For our bimodal galaxies that show separated sequences in the metal-poor region, the end of bursty star formation also sets an important timescale, but for a different reason: during the bursty phase, the accreted gas is not sufficiently metal-poor due to enrichment by outflows.

While the appearance of $\alpha$-bimodality has been extensively studied throughout the Milky Way disk, efforts to find similar abundance patterns in nearby spirals are in their earliest stages. Recent results from JWST NIRSpec suggest that M31 doesn't show bimodality at the analog of the solar radius (\citealp{Nidever_M31}; however see \citealp{Kobayashi_M31} for an alternative interpretation); LMC also doesn't exhibit any signs of bimodal patterns \citep{Nidever2020}. Does that mean that Milky Way abundance trends are unique? Among 11 Milky Way-like galaxies in FIRE-2 only three do not show any signatures of bimodality and the rest have more or less prominent signatures of double (or multiple in the case of m12b) sequences, which suggests that the presence of a low-$\alpha$ sequence is not a rare feature. However the separation between sequences in simulated galaxies is smaller than in the Milky Way, and detection of such features would be challenging given the present-day observational abundance uncertainties \citep{Patel22}.
\citet{Yu21} revealed that in FIRE-2 simulations, galaxies with longer phases of steady star formation typically have higher thin-disk fractions. 
M31 is more thick-disk dominated than the Milky Way \citep{Dalcanton2023}, which could mean that the steady phase of star formation in M31 has been shorter than in our Galaxy so it didn't have enough time to develop a strong low-$\alpha$ sequence; note, however, that most of the FIRE-2 galaxies formed a low-$\alpha$ sequence much later than the Milky Way. 
Also it is interesting to note, that in one of our Local Group-like pairs, \texttt{Thelma}-\texttt{Louise}, the more massive host has a very weak low-$\alpha$ sequence, while the less massive host displays a very prominent bimodality.

\section{Conclusions}\label{sum}
In this work, we have analyzed 11 Milky Way-mass galaxies from the FIRE-2 simulations and studied the distribution of their disk stars in chemical abundance space. Five galaxies show clear $\alpha$-bimodality on the [Fe/H] - [Mg/Fe] plane in the form of two sequences separated in the low-metallicity regime; in two galaxies the sequences converge at around solar [Fe/H]. The high-$\alpha$ population is older, kinematically hotter, and more centrally concentrated, in qualitative agreement with trends observed in the Milky Way.

Our analysis revealed key differences in the gas accretion histories of bimodal versus non-bimodal galaxies. While their early accretion histories are similar, over the last 5 -- 6 Gyrs bimodal galaxies experience increasing inflow of relatively metal-poor gas, compared to the decreasing (and generally more metal-rich) gas inflow in non-bimodal systems. The accreted gas has relatively high angular momentum and settles at large radii, resulting in the notable growth of the gaseous disk in bimodal galaxies. Star formation proceeds in the ISM diluted by this metal-poor gas inflow and results in the formation of a low-$\alpha$ sequence with a metal-poor tail.

A case study of galaxy \texttt{m12f} highlights that gas-rich mergers are able to produce different bimodal pattern than the one resulting from smooth accretion from CGM. However, in this case (and in the case of three bimodal galaxies that show bimodality at solar and super-solar metallicities, \texttt{Romulus}, \texttt{Romeo}, \texttt{Juliet}) dilution is only one of the factors at play and the behaviour of star formation history should be considered (see Barry et al. in prep.)

\section*{Acknowledgements}
The authors would like to thank the reviewer for providing helpful and constructive comments. HP also thanks Artyom Tanashkin for encouraging and useful discussions. 
AW and MB received support from: NSF via CAREER award AST-2045928 and grant AST-2107772; HST grant GO-16273 from STScI.

FIRE-2 simulations were generated using: XSEDE, supported by NSF grant ACI-1548562; Blue Waters, supported by the NSF; Frontera allocations AST21010 and AST20016, supported by the NSF and TACC; Pleiades, via the NASA HEC program through the NAS Division at Ames Research Center. We acknowledge the Texas Advanced Computing Center (TACC) at The University of Texas at Austin for providing computing resources that contributed to our results.

\section*{Data availability}
The FIRE-2 simulations are publicly available \citep[][]{wetzel2022} at \url{http://flathub.
flatironinstitute.org/fire}. Additional FIRE simulation
data is available at \url{https://fire.northwestern.edu/data}.
A public version of the GIZMO code is available at \url{http:
//www.tapir.caltech.edu/~phopkins/Site/GIZMO.html}.



\bibliographystyle{mnras}
\bibliography{example} 




\appendix

\section{Distribution of [Mg/Fe] in bins by [Fe/H]}
\begin{figure*}
    \centering
    \includegraphics[width=\textwidth]{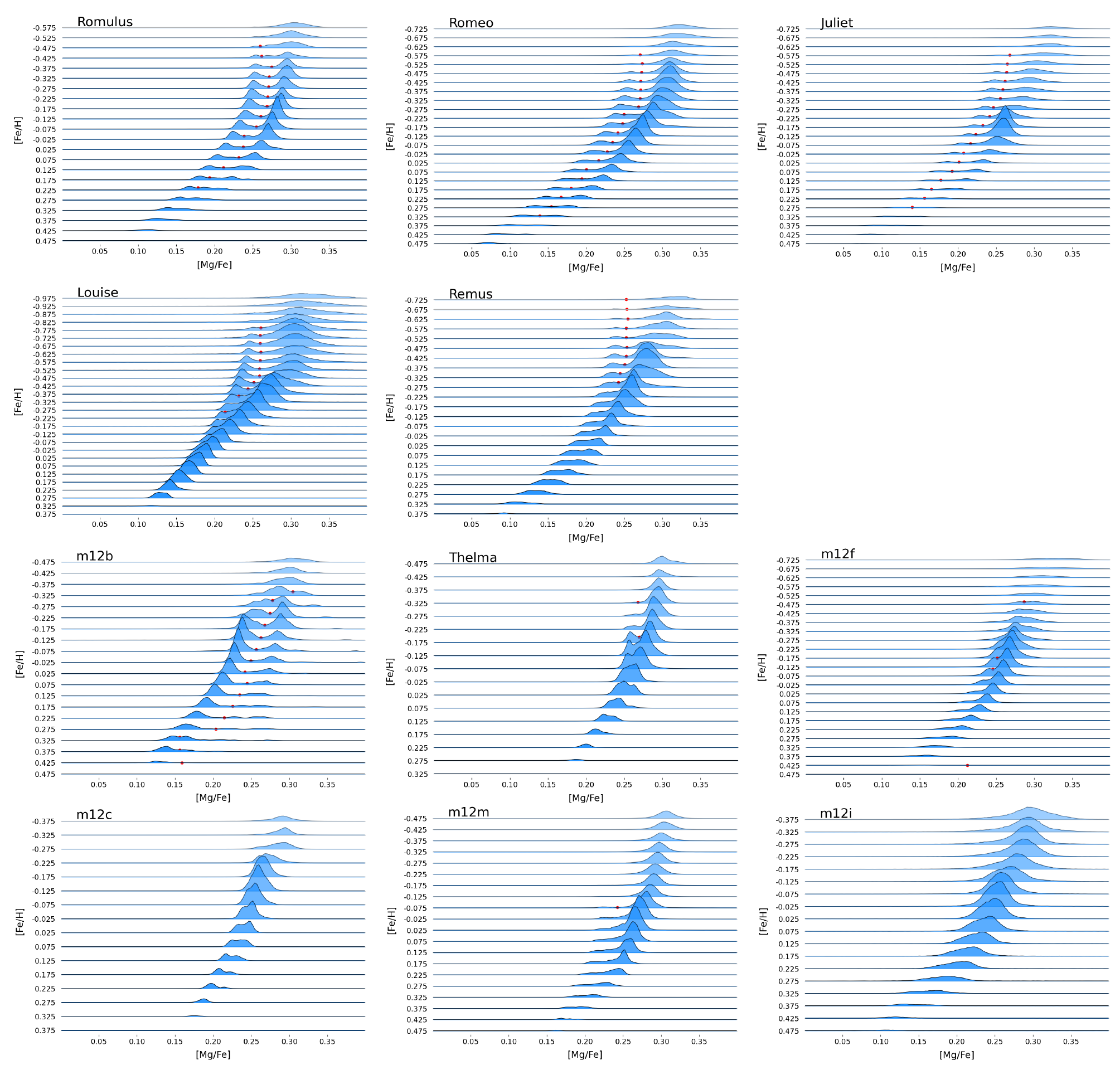}
    \caption{Distributions of $\mathrm{[Mg/Fe]}$ for the disk stars ($j_z/j_{\mathrm{circ}} > 0.2$) binned by $\mathrm{[Fe/H]}$, red dots mark the position of the minimum when two peaks are detected.}
    \label{fig:all_joy}
\end{figure*}

In order to classify galaxies based on the prominence of their bimodal features, we examine the distributions of [Mg/Fe] in narrow ($\Delta\mathrm{[Fe/H]} = 0.05$) metallicity bins, which are shown on Fig.~\ref{fig:all_joy} for each galaxy in our sample. Red dots mark the location of minima when double peaks in [Mg/Fe] distribution are detected. To locate peaks we applied \texttt{scipy} function \texttt{find\_peaks} with prominence of 0.03 and distance (minimal separation between peaks) of 0.025 to the normalized distribution. Galaxies \texttt{Louise}, \texttt{Remus}, \texttt{Romulus}, \texttt{Romeo}, and \texttt{Juliet} are labeled as ``bimodal'', \texttt{Thelma} and \texttt{m12f} -- as ``weakly-bimodal'' as the peaks could be identified in those galaxies with a lower threshold on distance and over a short range of metallicities. Galaxies \texttt{m12m}, \texttt{m12i}, and \texttt{m12c} do not show bimodality, and galaxy \texttt{m12b} present an extreme case of strong bimodality and is examined separately in Barry et al (in prep.)

\section{Evolution of median [Fe/H] and [Mg/Fe] in radial bins}
\begin{figure*}
    \centering
    \includegraphics[width=\textwidth]{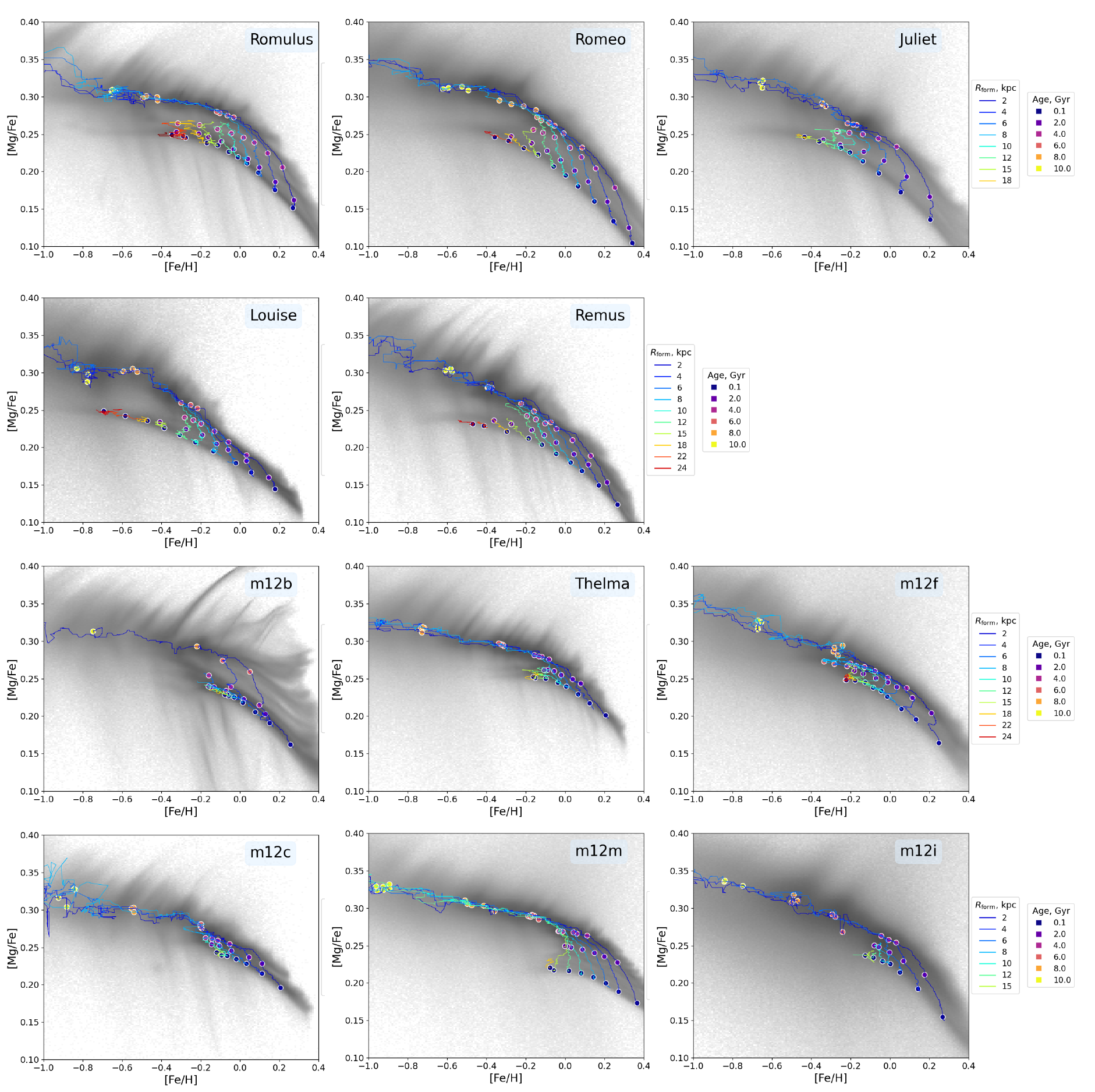}
    \caption{Elemental evolution tracks for stars binned by birth radii. The dots mark the age of the stars with corresponding median [Fe/H] and [Mg/Fe]. The greyscale histogram in the background shows the total distribution of stars formed within $R_{\ast,~98}(z=0)$.}
    \label{fig:all_tracks}
\end{figure*}

Fig.~\ref{fig:all_tracks} shows the evolutionary tracks of the median [Fe/H] and [Mg/Fe] for stars born within 0.1 kpc-wide radial bins for all galaxies in our sample.  The dots on Fig.~\ref{fig:all_tracks} mark the age of the stars born at each radius with given median [Fe/H] and [Mg/Fe]. The histogram in the background shows the distribution for all stars with formation radius smaller than $R_{\ast,~98}$ at z=0.

\bsp	
\label{lastpage}
\end{document}